\documentclass[a4paper,12pt]{article}

\usepackage{apacite}
\usepackage{xcolor}
\usepackage{url} 
\usepackage{lineno}

\usepackage{amsmath, amssymb, graphicx,color}
\usepackage{natbib}

\usepackage[top=3cm, bottom=3cm, left=3cm, right=3cm]{geometry}
\definecolor{webblue}{rgb}{0, 0, 0.5} 
\usepackage[pdfpagelabels,  
plainpages=true,
bookmarks=true,
bookmarksnumbered=true,
hypertexnames=true,
breaklinks=true,%
linkcolor=webblue,]{hyperref}
\hypersetup{
colorlinks  = true,
linkcolor = webblue,
urlcolor = webblue,
citecolor = webblue,
pdfauthor   = {John F. Rudge},
pdfsubject = {},
pdftitle    = {Convection},
pdfkeywords = {},
}

\newcommand{\D}{\mathrm{D}}
\renewcommand{\d}{\mathrm{d}}
\newcommand{\e}{\mathrm{e}}

\newcommand {\pdd}[2]{\frac{\partial #1}{\partial #2}}

\begin{document}
\title{Gravity, topography, and melt generation rates from simple 3D models of 
mantle convection}
\author{Matthew E. Lees\thanks{Current address: School of Earth, 
Energy, and Environmental Sciences, Stanford University, CA, USA.}, 
John F. Rudge, and Dan 
McKenzie\\
\small{Bullard Laboratories, Department of Earth Sciences,University of 
Cambridge}}
\maketitle

\begin{abstract}
Convection in fluid layers at high Rayleigh number (Ra $\sim 10^6$) have a 
spoke pattern planform. Instabilities in the bottom thermal boundary layer 
develop into hot rising sheets of fluid, with a component of radial flow 
towards 
a central upwelling plume. The sheets form the ``spokes'' of the pattern, and 
the plumes the ``hubs''. Such a pattern of flow is expected to occur beneath 
plate interiors on Earth, but it remains a challenge to use observations to 
place constraints on the convective planform of the mantle. Here we present 
predictions of key surface observables (gravity, topography, and rates of melt 
generation) from simple 3D numerical models of convection in a fluid layer. 
These models demonstrate that gravity and topography have only 
limited sensitivity to the spokes, and mostly reflect the hubs (the rising and 
sinking plumes). By contrast, patterns of melt generation are more 
sensitive to short wavelength features in the flow. There is the potential 
to have melt generation along the spokes, but at a rate which is relatively 
small compared with that at the hubs. Such melting of spokes can only occur 
when the lithosphere is sufficiently thin ($\lesssim 80$ km) and mantle water 
contents are sufficiently high ($\gtrsim 100$ ppm). The distribution of 
volcanism across the Middle East, Arabia and Africa north of equator suggests 
that it results from such spoke pattern convection.
\end{abstract}

\section{Introduction}

What is the planform of mantle convection? 
The largest, and most obvious, planform in the convection system is 
that associated with plate motions, which involves horizontal scales 
as large as 10,000 km (e.g. the Pacific plate). However, it is also clear 
that convection takes place at shorter horizontal scales. 
This scale of convection manifests in hot-spot volcanism, and the 
swells and troughs in gravity and topography observed at wavelengths of around 
1,000 to 2,000 km \citep{McKenzie1994,Crosby2006,Crosby2009}. A well-cited 
example of this comes from Africa \citep{Holmes1965,McKenzie1975,
Burke1996,Jones2012}, which shows a clear pattern of swells and troughs across 
the continent. The strong correlation between the patterns of gravity and 
topography, and in particular the characteristic ratio of around 50 
mgal~km$^{-1}$ between the two, has been used to strongly argue for convective 
support of the topography \citep{McKenzie1994,Crosby2006,Jones2012}. 

In theory, maps of gravity and topography should provide 
information on the planform of mantle convection. But extracting that 
information is challenging. While gravity provides information on density 
variations within the Earth, the process of inverting gravity data for density 
is highly non-unique. However, density variations within the Earth on 
length scales of hundreds of kilometers are not 
arbitrary, but are controlled by the fluid dynamics of convection. 

The aim of the present manuscript is to get a better understanding of 
the surface expressions of mantle convection from a series of 
the simplest possible 
numerical simulations of convection, and to compare these with geophysical and 
geological observations.  The fluid dynamical problem  
is one that has been extensively studied and discussed in the literature: 
Rayleigh-B\'{e}nard thermal convection in a 3D rectangular box with a fluid of 
constant viscosity at Rayleigh numbers around $10^6$. This problem was first 
studied 
using laboratory experiments in large aspect ratio tanks, using a layer of 
silicone oil whose depth was a few 
centimetres 
\protect{\citep{
Busse1971,Busse1974,Richter1975}}.  These experiments showed that the planform 
of 
the 
convection was a spoke pattern, with hot rising plumes joined to each other by hot sheets 
near the lower thermal boundary layer, and cold sinking plumes by colds sheets 
near the top thermal boundary layer.  \citet{White1988} carried out similar 
experiments 
using a 
fluid whose viscosity was a strong function of temperature, and showed that 
at high Rayleigh number the convective planform was also a spoke pattern.

It is now possible to carry out three-dimensional time-dependent 
fully-resolved numerical experiments at Rayleigh numbers of $10^5 - 10^7$, 
and the results of many such experiments have been reported 
\citep{Houseman1990,Christensen1991,Weinstein1991,Tackley1993,
Tackley1996,Larsen1997,Sotin1999,Zhong2005,Galsa2007,Vilella2017,Vilella2018}. 
Such experiments have an important advantage over tank experiments, because it 
is straightforward to calculate the geophysical observables from the numerical 
solutions.

We carried out our calculations with an infinite Prandtl number 
fluid, and used the 
Boussinesq approximation throughout.  We used large aspect ratio boxes, 
similar to those previously used in tank experiments, to allow the convective 
circulation to determine its own planform rather than being dominated by the 
lateral boundaries. We use our experiments to show which features of the 
observations are readily explained by the simplest models, and which features 
are not. We do not attempt to construct a realistic model of 
the Earth.  In this respect our aim is the same as that of the early tank 
experiments, and, in addition, to extend them to encompass the observables: 
the gravity field, surface deformation and melt generation. In particular, by 
allowing the density variations to arise naturally from fluid dynamics, such 
experiments allow an exploration of short wavelength ($ < 100$ km) temperature 
variations, which are likely to be most clearly expressed by volcanism. 

Two effects that are known to be important in the Earth are 
not taken into account in the simple constant viscosity Boussinesq model we 
use.  The first is the variation of viscosity with temperature.  It is this 
effect that produces plates, and therefore our modelling does not include the 
dynamics of plate motions.  The other effect is viscous dissipation, which is 
intimately related to vertical density variations that result from lithostatic 
pressure \protect{\citep{Spiegel1960,Jarvis1980,Schubert2001a}}.  Viscous heat 
generation has little effect on the circulation even when 
the relevant term in the equations cannot be neglected.  Such heating occurs 
in boundary layers where temperature gradients are large.  As a result the 
entropy and potential temperature (appendix \ref{sec:ptemp}) is little affected 
\protect{\citep{Jarvis1980}}, though the convection becomes more time 
dependent.

The approach taken here complements a popular alternative approach to 
modelling gravity and topography using information from seismic tomography 
\citep{Hager1989,Flament2013}. In such studies, estimates of density 
variations within the mantle are inferred from tomography and used to make 
predictions of gravity and dynamic topography. Though these studies have had  
some success at predicting the very long wavelength ($>$ 6,000 km) features of 
Earth's gravity, they depend on knowing the relationship between density $\rho$ 
and 
the seismic velocities $V_P$ and $V_S$, and also on a rheological model of the mantle.  Furthermore, since these calculations are based on seismic 
tomography, they are limited by its resolution, which is not yet sufficient to 
map rising and sinking plumes in the upper mantle. Our approach also departs 
from the common assumption of many Earth Scientists, who believe the 
convective planform of mantle convection consists solely of plumes and 
the plates.  This assumption 
arises from the work of \citet{Wilson1963} and 
\citet{Morgan1971}, who showed that the 
relative motion between major volcanic centres beneath plate interiors was 
sufficiently slow that they could be used to define a single world-wide 
reference frame.  Their ideas have been enormously influential. But they are 
based on an intuitive conception about the planform of high Rayleigh number 
convection, rather than on fluid dynamical experiments.  They also predate our 
understanding of polybaric melt generation. 

The manuscript is organized as 
follows. Section \ref{sec:numeric} describes the fluid dynamical simulations, 
and how they are scaled to parameter values appropriate for the Earth's 
mantle. Section \ref{sec:gravtopo} discusses the predicted gravity, topography 
and their spectral properties. Section \ref{sec:melt} discusses melt 
generation. Section \ref{sec:obs} compares the results from the fluid dynamical 
experiments with the observed gravity and topography, and with the volcanism 
of Africa and the Middle East, and conclusions follow in section 
\ref{sec:conclude}. An appendix provides further technical details on the 
simulations and data processing.

\section{Numerical experiments}\label{sec:numeric}

We ran 12 numerical experiments of isoviscous thermal convection in a 
rectangular box. Temperature was fixed at the top and bottom boundaries. To 
examine the influence of dynamical boundary conditions, runs were made for all 
combinations of freely slipping or rigid boundary 
conditions on the top and bottom boundaries. Reflection boundary conditions were applied at the side boundaries. 

All convection simulations were performed using v2.01 of the ASPECT mantle 
convection code \protect{\citep{Dannberg2016a,Heister2017,Bangerth2018}}. The 
code was 
used to solve the dimensionless versions of the Boussinesq governing equations 
of thermal convection in an 
$8 \times 8 \times 1$ rectangular box, through a small modification of the 
``convection-box'' example discussed in the ASPECT manual. The governing 
equations in dimensional form are
\begin{linenomath}
\begin{gather}
\nabla \cdot \mathbf{v} = 0, \\
- \nabla \mathcal{P} + \eta \nabla^2 \mathbf{v} = - \rho_0 g \alpha \theta 
\hat{\mathbf{z}} , \label{eq:stokes} \\
\frac{\D \theta}{\D t} = \kappa \nabla^2 \theta,
\end{gather}
\end{linenomath}
where $\mathbf{v}$ is the velocity, $\mathcal{P}$ is the difference in pressure 
from hydrostatic,  $\theta$ is potential temperature, $\rho_0$ is the reference mantle density, $\alpha$ is the thermal 
expansivity, $\eta$ is the viscosity, and 
$\kappa$ is the thermal diffusivity. We assume constant thermal conductivity, 
constant heat capacity, constant viscosity, and constant thermal expansivity. 
In dimensionless form the governing equation are
\begin{linenomath}
\begin{gather}
\nabla \cdot \mathbf{v} = 0, \\
- \nabla \mathcal{P} +  \nabla^2 \mathbf{v} = - 
\text{Ra} \theta \hat{\mathbf{z}} , \\
\frac{\D \theta}{\D t} = \nabla^2 \theta, \label{eq:dimless_energy_cons}
\end{gather}
\end{linenomath}
where all lengths have been scaled by the layer depth $d$, time by the diffusion time $d^2/\kappa$, pressure by $\eta \kappa/d^2$, and potential temperature by the potential temperature difference $\Delta T_p$ across 
the layer. Just 
one dimensionless parameter describes this simple system, the Rayleigh 
number, defined by
\begin{linenomath}
\begin{equation}
 \text{Ra} = \frac{\rho_0 g \alpha \Delta T_p d^3}{\eta \kappa}.
\end{equation}
\end{linenomath}
Runs were performed at three different 
Rayleigh numbers: $\text{Ra}=10^5$, $3 \times 10^5$, and $10^6$. A uniform resolution of 32 cells in the vertical and 256 cells in the horizontal 
was specified for all simulations. Quadratic finite elements were used for 
temperature and velocity, and linear finite elements for pressure. Simulations 
were run until the system reached a quasi-steady state, which was monitored by 
examining the behaviour of the mean temperature and RMS velocity over time. 
Each simulation ran for a minimum of six times the thermal time constant for 
the layer ($=6 \, d^2/(\pi^2 \kappa)$).

\begin{figure}
 \includegraphics[width=\columnwidth]{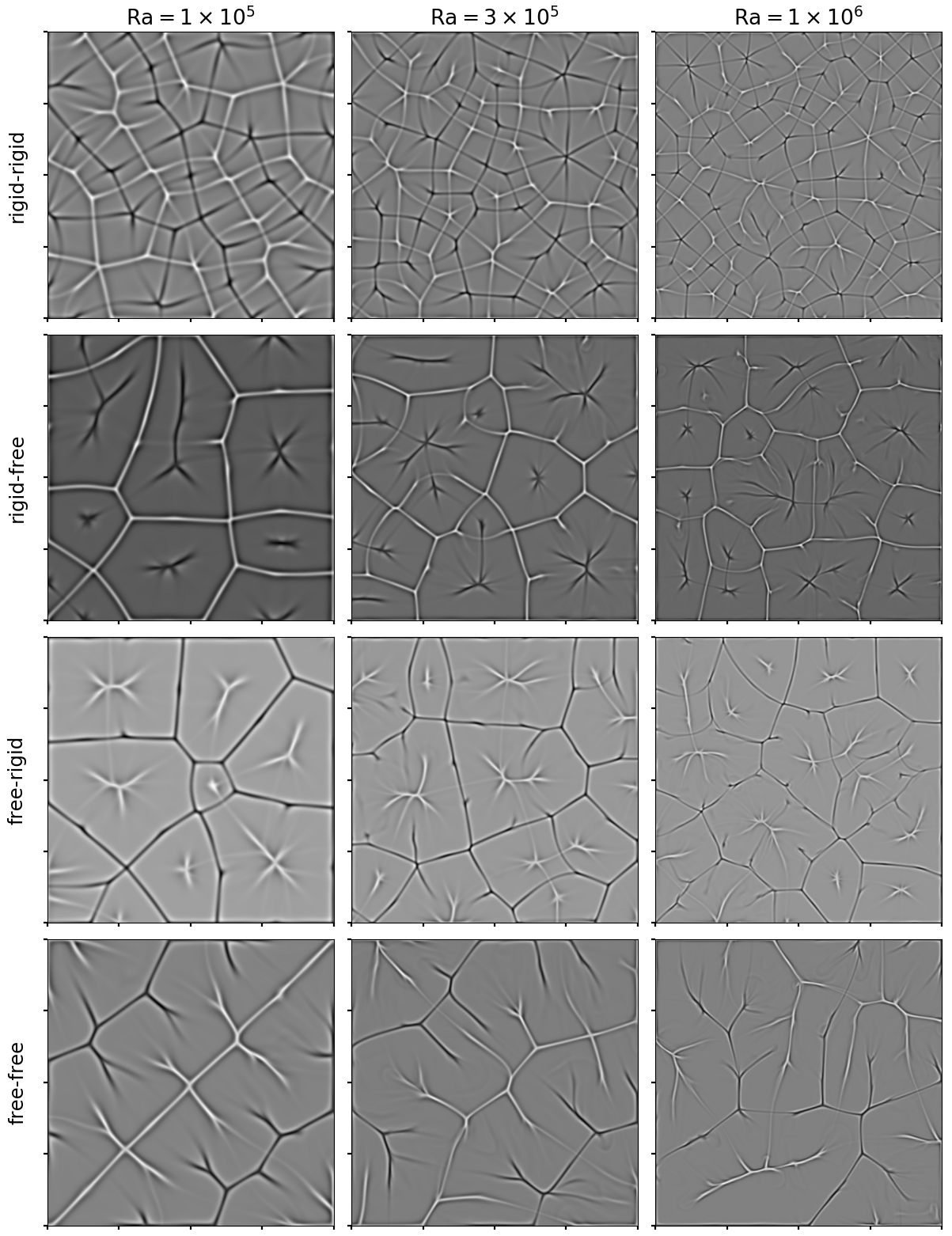}
\caption{Artificial shadowgraphs of the 12 numerical experiments of 
thermal convection in an $8 \times 8 \times 1$ rectangular box. Each column is 
at a given Rayleigh number, and each row is at a given choice of boundary 
conditions for the Stokes flow. In each case the first 
word corresponds to the top boundary, the second word to the bottom boundary; 
so free-rigid refers to a free-slip top and rigid bottom boundary 
condition.}
\label{fig:shadowgraph}
\end{figure}

Snapshots of the 12 experiments are shown in \autoref{fig:shadowgraph}. 
The images are made to mimic the shadowgraph visualization 
technique 
commonly used in laboratory experiments 
\protect{\citep{Busse1971,Busse1974,Richter1975,
Whitehead1977,White1988}}. To make a shadowgraph in the laboratory, light is 
shone from below through the 
layer of fluid and projected onto a screen. Refraction causes the light to 
focus and defocus according to the temperature variations within the fluid 
through which it 
passes. The effect is to make upwellings appear dark and downwellings 
appear bright. Mathematically, a shadowgraph produces a plot of the 
Laplacian of the vertically averaged temperature 
\citep{Jenkins1988,Travis1990b}, and this is how the images in 
\autoref{fig:shadowgraph} were computed.

The planform of convection at these Rayleigh numbers is spoke 
pattern \citep{Busse1974}. Hot and cold plumes develop from the top and bottom 
of the 
layer: the ``hubs" of the planform.  Hot and cold sheets, the ``spokes" of the 
planform, radiate from the plumes.  These sheets are formed by instabilities 
in the top and bottom boundary layers, and the shearing associated with the 
plumes suppresses instabilities with other geometries. The planform is 
time-dependent.


Several well-known features stand out from the shadowgraph pictures. The 
first is that as the Rayleigh number increases the thicknesses of the 
upwellings and downwellings become narrower. For the free-free and 
rigid-rigid cases the upwellings and downwellings are symmetric, but there 
is a notable asymmetry in the nature of upwellings and downwellings in the 
free-rigid and rigid-free cases \citep{Kvernvold1979,Weinstein1991}. Also 
notable is the change in horizontal distance between upwelling and 
downwelling with boundary condition: this distance is shorter for 
rigid-rigid simulations than for free-free simulations. 

\begin{table}
\begin{center}
 \begin{tabular}{lcc}
\hline
Quantity &  Symbol & Value \\ \hline
thermal expansivity & $\alpha$ & $4.0 \times 10^{-5}$ K$^{-1}$\\
acceleration due to gravity & $g$ & 9.81 m~s$^{-2}$\\
layer thickness & $d$ & 600 km\\
reference mantle density & $\rho_0$ & 3300 kg~m$^{-3}$\\
thermal conductivity & $k$ & 3.7 W~m$^{-1}$~K$^{-1}$\\
specific heat capacity & $C_p$ & $1.3 \times 10^3$ J~kg$^{-1}$~K$^{-1}$\\
\hline
 \end{tabular}
\end{center}
\caption{Common parameter values for all runs.  These values are appropriate 
for the top $\sim 100$ km of the upper mantle beneath the lithosphere}
\label{tab:params}
\end{table}

Our main interest here is the surface observables associated with mantle 
convection. The simulations were performed using dimensionless variables, 
with the dimensionless Rayleigh number as the only control parameter. To 
make predictions about observable quantities, these simulations must be 
scaled appropriately to Earth-like values. For most quantities we can 
simply use typical mantle estimates, and these values are given 
\autoref{tab:params}. We consider here just upper mantle convection, and 
so choose as an appropriate layer thickness $d=600$ km to scale all lengths.

Choosing an  appropriate scaling of temperature is less straightforward. 
The numerical 
simulations essentially provide 
dimensionless potential temperatures. 
The potential 
temperature is the temperature that the mantle material would have if it were 
moved to the Earth's surface isentropically and without melting 
(\citet{McKenzie1970}, appendix \ref{sec:ptemp}).

\begin{figure}
 \includegraphics[width=\columnwidth]
{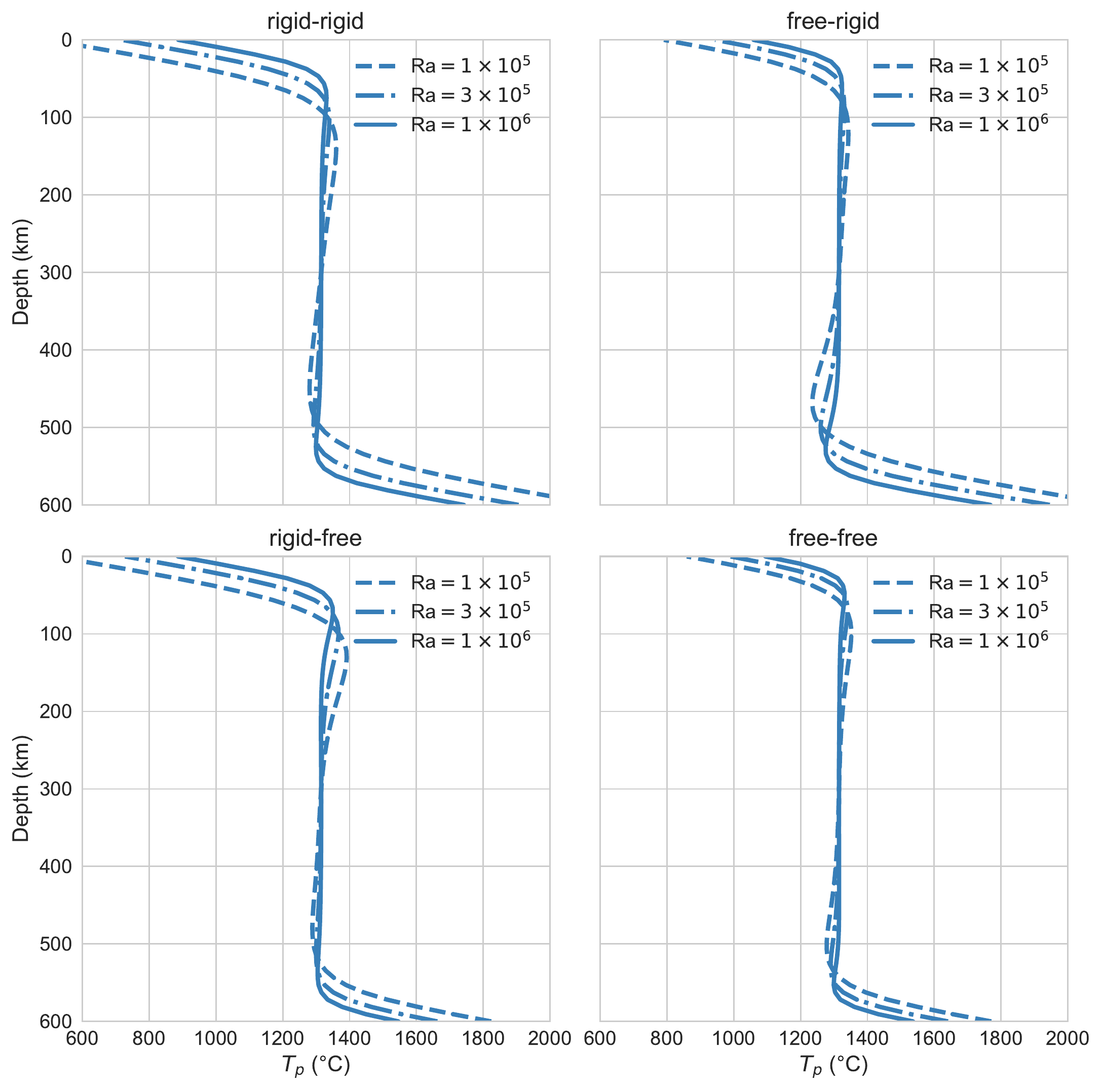}
\caption{Profiles of horizontally-averaged potential temperature. Zero 
depth represents the top of the convecting region (the base of the 
mechanical boundary layer). The four panels show the four choices of 
boundary condition. The dimensional scaling is such that the interior 
potential temperature is 1315$^\circ$C and the lithospheric thickness is 
100 km.}
\label{fig:temp_profiles}
\end{figure}

The scaling of the dimensionless 
temperature is required to satisfy two conditions: 
The first is that the 
average interior temperature must correspond to a mantle potential temperature 
of 
1315$^\circ$C. This choice ensures that the thickness of the oceanic crust is 
7 km, generated by isentropic decompression beneath a spreading ridge using  
the parametrisation of \citet{Katz2003}. The second 
condition arises from the top of the convecting system 
not being at the Earth's surface. We envisage that there is 
a rigid mechanical boundary layer (MBL) which 
separates the Earth's surface from the top of the convecting system. 
The temperature near the base of the MBL is where $T/T_s$, 
where $T_s$ is the melting temperature, is highest, and therefore the 
viscosity is lowest \protect{\citep[e.g.][]{Frost1982}}.  For 
this reason we used both 
free-slip and rigid boundary conditions on the top surface of the 
convecting 
box.  A stress-free boundary is probably the better approximation to the 
behaviour of the real Earth. 
Heat transfer through the MBL is purely by conduction, and 
we assume the thermal profile is linear through this region, and equal to 
$0^\circ$C at the Earth's surface. At the top of the convecting region we 
assume that both the horizontally-averaged temperature and heat flux are 
continuous with that in the MBL. Finally we determine the thickness of 
the MBL by prescribing a lithospheric thickness, which we define as the 
intersection of the linear conductive profile of the MBL with an isentropic 
profile at the interior mantle potential temperature. In what follows the 
lithosphere thickness is fixed at 100 km, except when considering melt 
generation where we have varied this parameter, as melt 
generation is particularly sensitive to it. This choice of interior 
potential temperature and lithospheric thickness fixes the heat flux for all 
simulations to be 50 mW~m$^{-2}$, which is similar to that through old sea 
floor \citep{Hasterok2013}. 
The resulting associated parameters, which include the upper mantle viscosity, 
are given in \autoref{tab:params2}.
The horizontally-averaged potential temperature profiles after scaling 
are shown in \autoref{fig:temp_profiles}.

\begin{table}
\begin{center}
 \begin{tabular}{ccccccc}
\hline
BC&Ra&$\eta$ (Pa s)&MBL (km)&$\Delta T_p$ (K) & $h_\text{RMS}$ (m) &$\Delta 
g_\text{RMS}$ (mgal)\\
\hline
rigid-rigid&$10^6$&$2.8\times 10^{20}$&66.8&855 &307&16.3\\ 
rigid-rigid&$3\times 10^5$&$1.3 \times 10^{21}$&53.8&1188 &636&32.9\\ 
rigid-rigid&$10^5$&$5.3\times 10^{21}$&37.2&1648 &1300&63.7\\ 
free-rigid&$10^6$&$2.3 \times 10^{20}$&80.1&709 &244&11.2\\ 
free-rigid&$3\times 10^5$&$1.1\times 10^{21}$&71.2&1002 &495&21.0\\ 
free-rigid&$10^5$&$4.4 \times 10^{21}$&59.4&1349 &958&37.1\\ 
rigid-free&$10^6$&$2.1 \times 10^{20}$&66.7&660 &325&15.5\\ 
rigid-free&$3\times 10^5$&$1.0 \times 10^{21}$&54.6&935 &672&32.6\\ 
rigid-free&$10^5$&$4.2\times 10^{21}$&38.6&1307 &1333&65.2\\ 
free-free&$10^6$&$1.4 \times 10^{20}$&82.8&442 &235&10.1\\ 
free-free&$3\times 10^5$& $7.0 \times 10^{20}$ &74.7&651 &476&19.8\\ 
free-free&$10^5$&$3.0\times 10^{21}$&64.6&912 &943&37.7\\ 
\hline
\end{tabular}
\end{center}
\caption{Parameter values and magnitudes of observables for each simulation 
which yield a lithospheric 
thickness of 100~km and an interior potential temperature of 
1315$^\circ$C. Columns from left to right: BC, boundary conditions at 
top-bottom; Ra, the Rayleigh number; $\eta$, viscosity; MBL, thickness of 
the mechanical boundary layer; $\Delta T_p$, potential temperature 
difference across the fluid layer; $h_\text{RMS}$, root mean square of the 
dynamic topography at the top of the convecting region (as plotted in 
\autoref{fig:topo_nofilter}); $\Delta 
g_\text{RMS}$, root mean square of the gravity anomaly at the top of the 
convecting region (as plotted in 
\autoref{fig:grav_nofilter}).}
\label{tab:params2}
\end{table}

With Rayleigh numbers in the range $10^5$ to $10^6$, \autoref{tab:params2} 
shows mantle viscosity values vary from $1.4 \times 10^{20}$~Pa~s to $5.3 
\times 10^{21}$~Pa~s, which are around the range 
expected for the upper mantle ($\sim10^{21}$~Pa~s). The higher the Rayleigh 
number, the lower the inferred viscosity, the smaller the potential temperature 
difference across the layer, and the thicker the MBL. Appendix 
\ref{sec:Ra_scaling} discusses further the behaviour of the parameters with 
Rayleigh number.

\section{Gravity and topography}\label{sec:gravtopo}

Examples of the gravity and topography at the top of the convecting box 
when there is no lithosphere present 
are shown in Figures \ref{fig:grav_nofilter} and \ref{fig:topo_nofilter}, 
calculated from the expressions in \citet{Parsons1983} 
(appendix \protect{\ref{sec:flexfilter}). The gravity 
field takes account of the contribution from the thermal expansion 
of the fluid and that from the topography (assuming deformable top and 
bottom boundaries).  
Both shadowgraphs and the plots of gravity and topography are filtered 
versions of the temperature field in the box, but they involve distinctly 
different kinds of filters.  Since the shadowgraph represents the Laplacian of 
the vertically-averaged temperature, it is equally sensitive to temperature 
variations at all depths, and acts to emphasise short-wavelength features. By 
contrast, gravity and topography act to attenuate short-wavelength features in 
the temperature field, particularly those at depth \citep{Parsons1983}. 

Before comparing the results from the numerical models  with the observations 
a further filter needs to be be applied, to account for the effect of the 
overlying mechanical boundary layer. This layer has a number of effects. 
First, gravity anomalies are attenuated by the thickness of the layer. Second, 
the elastic properties of the overlying plate acts to filter out the 
topographic expression of short-wavelength features, with the magnitude of 
this effect dependent on the assumed effective elastic thickness ($T_e$). 
Figures \ref{fig:grav_filter} and \ref{fig:topo_filter} show the expected 
gravity and topography at the surface, after applying a filter for the 
mechanical boundary layer, assuming an elastic thickness $T_e = 30$~km. The 
effect of all this filtering is to remove much of the short wavelength 
information that is visible in the shadowgraph images. In particular, the 
filters emphasise the ``hubs'' of the convective 
pattern, and suppress the expression of the ``spokes''. As will be seen in the 
next section, the effect of the 
filter on melt generation is even more important.  As the thickness of the MBL 
increases it first suppresses melt generation in the spokes and then in the 
hubs.

\begin{figure}
\includegraphics[width=\columnwidth]
{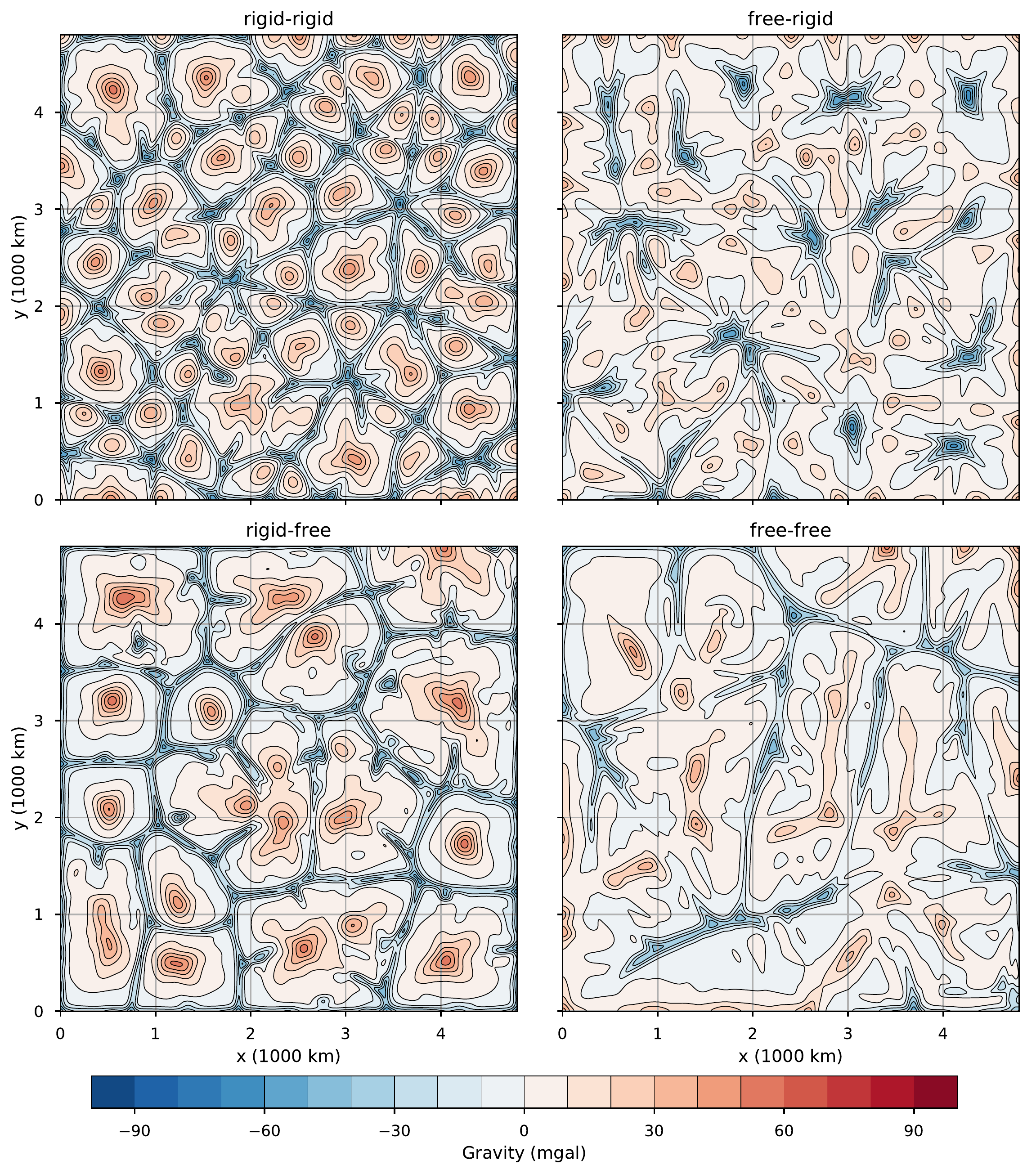}
\caption{Gravity anomalies at the top of the convecting box for the 
$\text{Ra}=10^6$ simulations. 
The region shown is approximately the same size 
as the region outlined by the thick black line in \autoref{fig:Fig_16}.
Each panel 
shows a different combination of boundary conditions. In each case the first 
word corresponds to the top boundary, the second word to the bottom boundary; 
so free-rigid refers to a free-slip top and rigid bottom boundary 
condition.}
\label{fig:grav_nofilter}
\end{figure}

\begin{figure}
\includegraphics[width=\columnwidth]{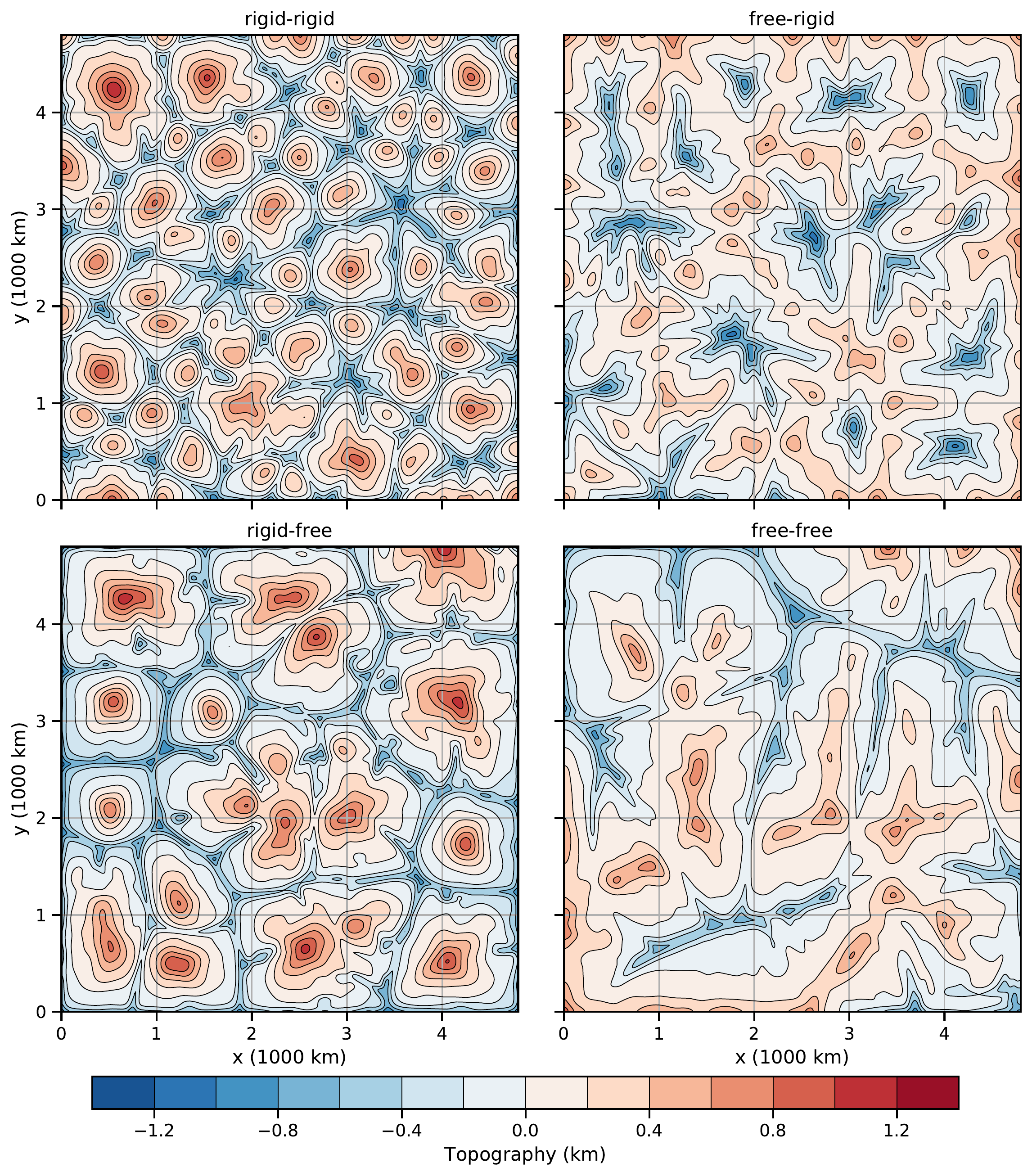}
\caption{Dynamic topography at the top of the convecting box for the 
$\text{Ra}=10^6$ simulations.}
\label{fig:topo_nofilter}
\end{figure}

\begin{figure} 
\includegraphics[width=\columnwidth]{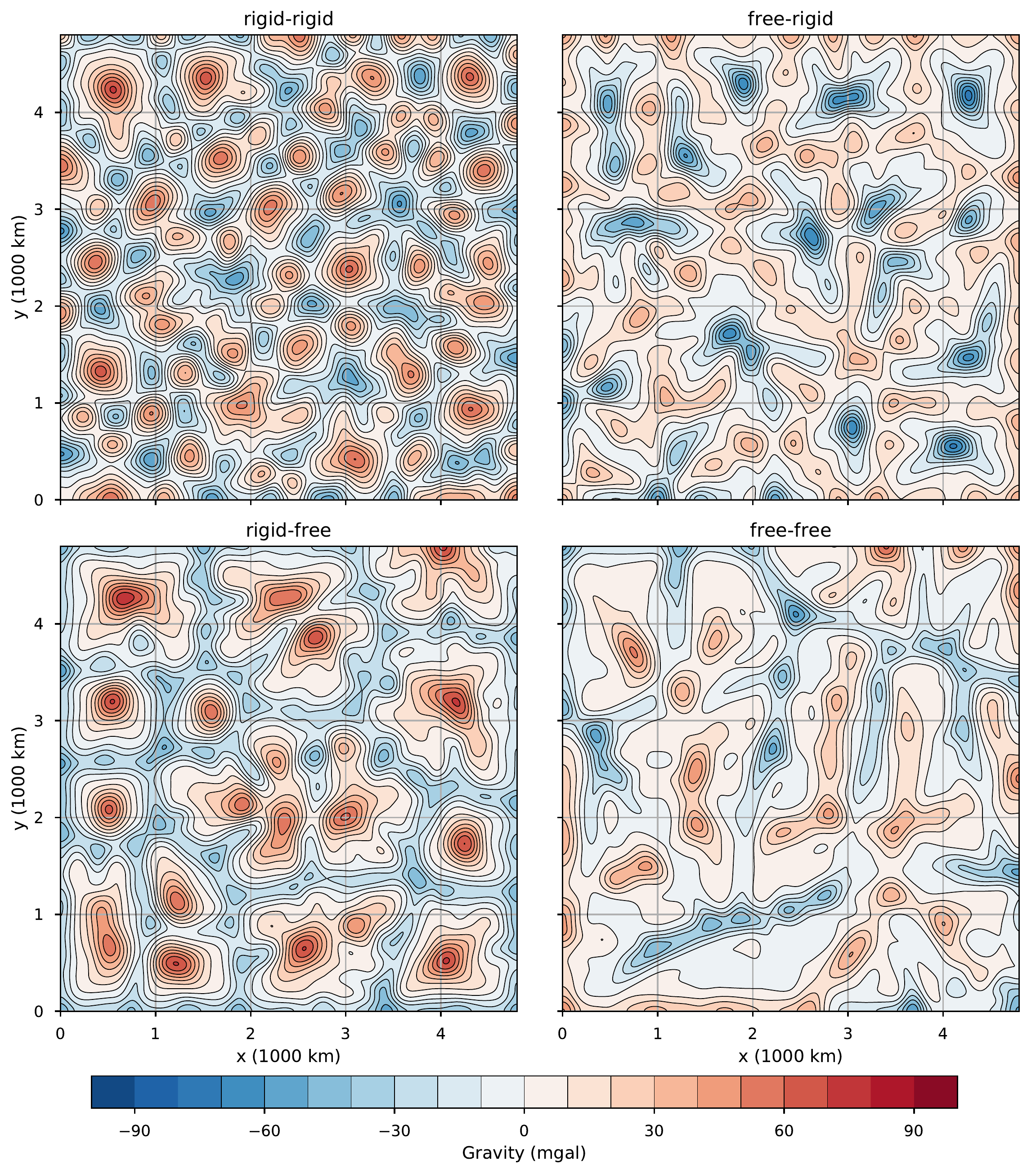}
\caption{Gravity anomalies expected at the Earth's surface when the 
lithospheric 
thickness is 100 km.  Plots are as in 
\autoref{fig:grav_nofilter}, except the attenuation of gravity anomalies 
through the mechanical boundary layer and an elastic plate with $T_e=30$~km 
has been taken into account.}
\label{fig:grav_filter}
\end{figure}

\begin{figure}
 \includegraphics[width=\columnwidth]{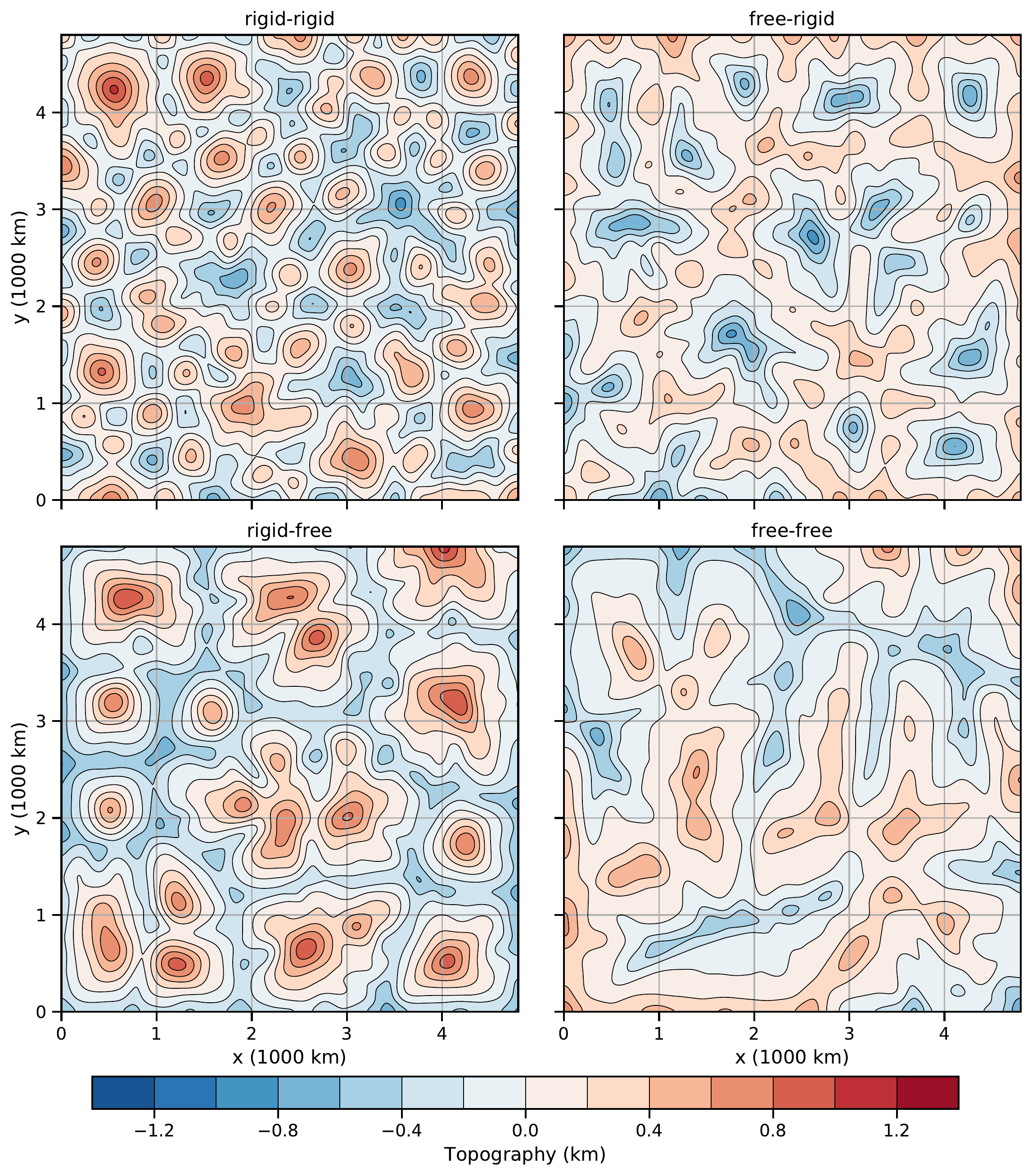}
\caption{Topography expected at the surface, after flexural filtering 
through an elastic plate with $T_e=30$~km.}
\label{fig:topo_filter}
\end{figure}

The images in Figures \ref{fig:grav_filter} and \ref{fig:topo_filter} 
demonstrate three key points: First, the predicted amplitude of gravity 
anomalies is comparable with the observed anomalies (compare scale of 
\autoref{fig:grav_filter} with \autoref{fig:Fig_18}(c)). Second, the 
spatial 
patterns of gravity anomalies and topography are sensitive to the assumed 
boundary conditions. Third, there is a good correlation between the 
topography and the gravity in  plots of the values of gravity and topography 
at each spatial  location (\autoref{fig:cross_plots}), which closely resemble 
similar plots that have been made using observed values of gravity and 
residual topography (e.g. see Figures 6 and 7 of \citet{Crosby2009}). The 
slope of these plots represents a characteristic average value of the 
admittance (ratio of gravity to topography) associated with the convection. A 
characteristic air-loaded admittance between 43-53 mgal~km$^{-1}$ is inferred 
from \autoref{fig:cross_plots}, corresponding to values of 30-37 mgal~km$^{-
1}$ when overlain by water.  These values are similar to those previously 
reported \citep{Parsons1983}. 

\begin{figure}
\includegraphics[width=\columnwidth]
{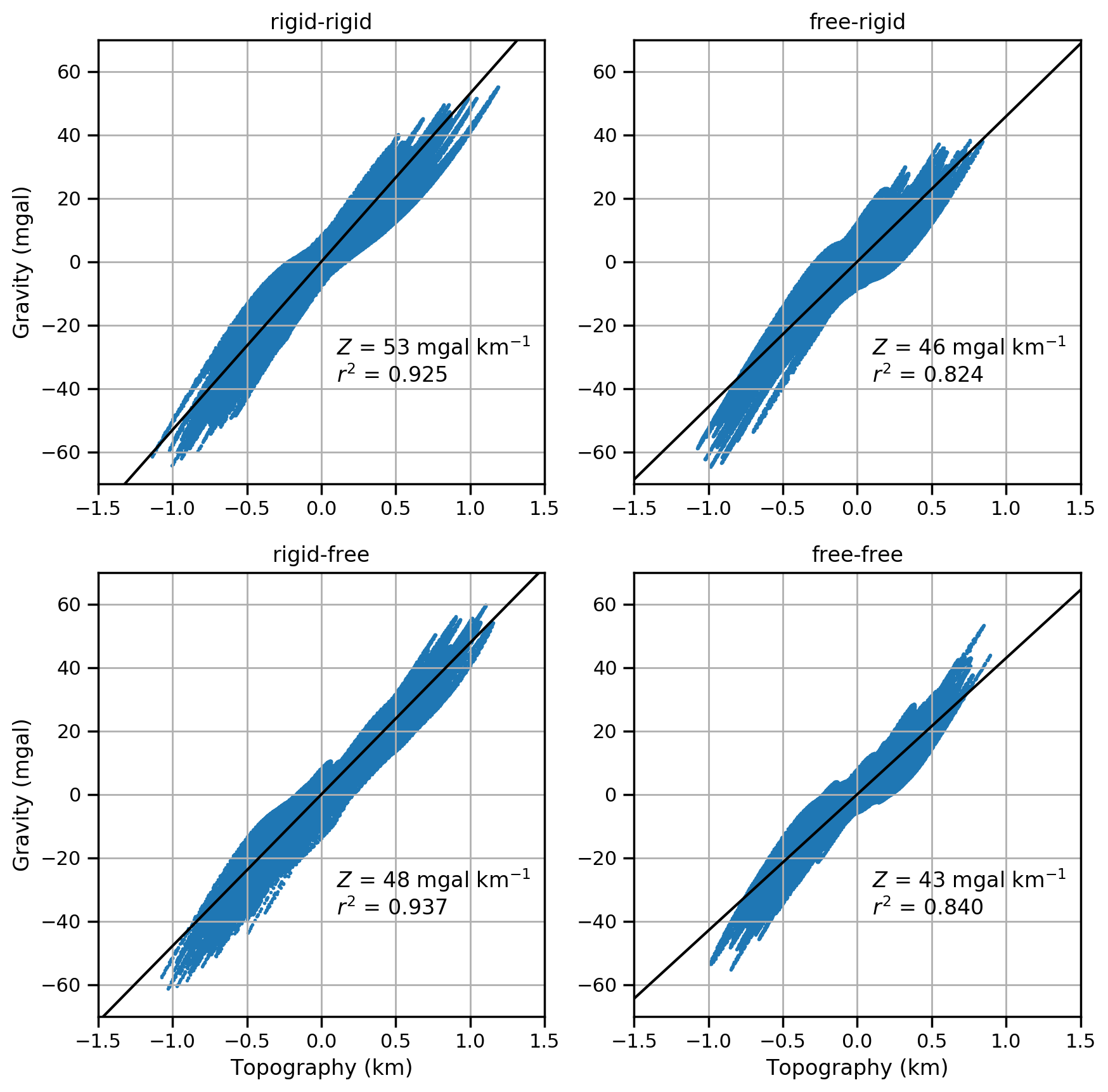} 
\caption{Cross plots of gravity against topography at the top of the 
convecting box for the four $\text{Ra}=10^6$ simulations. In each case a 
geometric regression line has been calculated, and marked on each plot is the 
slope $Z$ of that line, and $r^2$, the square of the correlation coefficient.
If the boxes are overlain by water the values of $Z$ are reduced by a factor 
of $(\rho_0-\rho_w)/\rho_0 \simeq 0.7$} 
\label{fig:cross_plots}
\end{figure}

\begin{center}
\begin{figure}
\includegraphics[width=0.99\columnwidth]{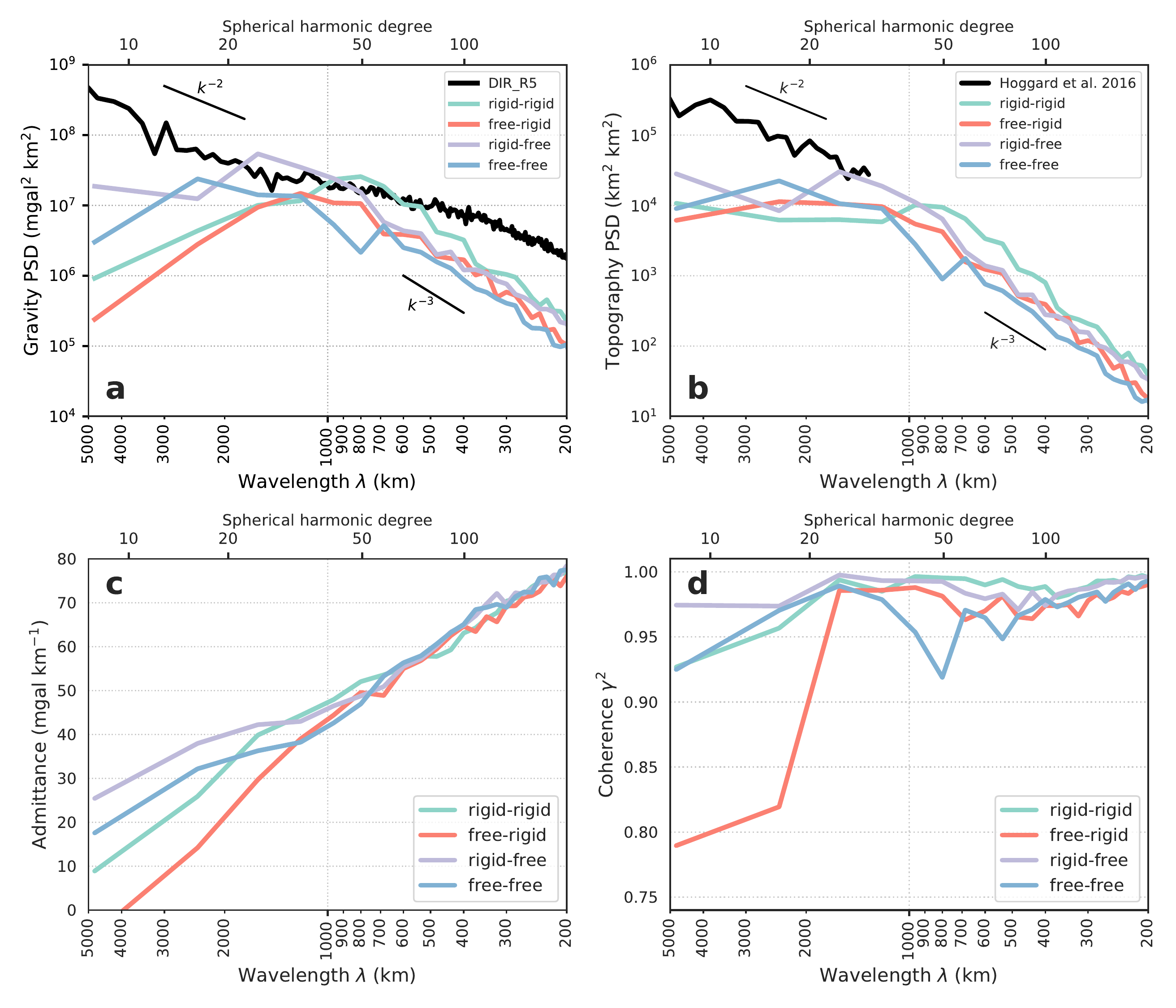}
\caption{(a) Power spectral density (PSD) of gravity anomalies at the top of 
the 
convecting region as a function of wave number $k=2 \pi / \lambda$, for the 
four 
convection simulations at $\text{Ra}=10^6$, along with that estimated for the 
Earth from the gravity model DIR-R5. Log scales are used for both axes. The 
approximate spherical harmonic degree estimated using Jeans relation is 
plotted along the top axis. Two thin line segments show slopes which yield a 
power law decay proportional to $k^{-2}$ (Kaula's rule) and $k^{-3}$. 
(b) Power spectral density of dynamic topography at the top of the 
convecting box as a function of wavenumber, 
along with that estimated for the Earth by \citet{Hoggard2016}. (c) 
Admittance (the ratio of gravity to topography) at the top of the convecting 
region as a function of wave number, assuming air-loading.  
(d) Coherence (square of the correlation between gravity and topography) at 
the top of the convecting region as a function of wave number.}
\label{fig:grav_spec}
\end{figure}
\end{center}

An alternative way to assess the predictions of gravity and topography is to 
work in the frequency domain rather than the spatial domain. 
\autoref{fig:grav_spec}(a) and (b) show 
the power spectral density of gravity and topography at the top of the 
convecting box (appendix \ref{sec:PSD}), along with estimates for the Earth. 
The figures show that the 
convection experiments have most power over a range of wavelengths from around 
500 to 2000 km, where the estimates of gravity anomalies are comparable in 
magnitude with those of the Earth. This behaviour reflects the fact that the 
convection organises into cells where the distance from one upwelling to the 
next is around twice the layer depth, corresponding to a wavelength around 
1200~km. Away from this broad peak the power from the convection experiments 
decays. At the short wavelengths shown in the plot, this decay in the power 
spectral density scales roughly as $k^{-3}$. The different boundary conditions 
in the convection experiments lead to subtle differences in the gravity 
spectra in \autoref{fig:grav_spec}. For example, there is greater power at 
long-wavelengths for those simulations with free-slip bottom 
boundary 
conditions than rigid. There is also greater power at short wavelengths for 
the simulations with rigid top boundary conditions than 
free-slip top boundary 
conditions. 

The power spectral density of gravity for the Earth is notably different from 
that of the convection experiments. Overall, the Earth's power spectral 
density decays broadly with wavenumber as $k^{-2}$, a power law which is 
referred to as Kaula's rule. In detail, the slope of decay flattens slightly 
in the 2000 km to 500 km wavelength band, before becoming steeper at 
wavelengths shorter than 500 km, but not decaying as steeply as in 
the convection experiments. The difference between the convection experiments 
and the Earth for wavelengths shorter than 500 km is to be expected: on Earth 
surface loading causes short wavelength topography that is supported by 
elastic stresses in the plate. The short wavelength power in the Earth's 
gravity field arises from surface loading, not mantle convection. At long 
wavelengths ($>$2000~km) mantle convection should play an important role in 
determining the gravity field on Earth, but the convection experiments here 
have notably less power than the Earth. Thus the upper mantle convection 
modelled here does not account for the magnitude of the long-wavelength 
portion of the Earth's gravity field. 

\autoref{fig:grav_spec}(b) shows the corresponding spectra for predictions of 
dynamic topography. The shape of the topography spectra of the convection runs 
is different from that for the gravity. The topography spectra are broadly flat 
up to around a wavelength of 1000 km, and then decay steeply at shorter 
wavelengths. Thus dynamic topography has relatively more long wavelength power 
than does gravity, and this can be seen in the space domain plots of 
\autoref{fig:grav_nofilter} and \autoref{fig:topo_nofilter} -- the topography 
plots look smoother than the gravity plots (as noted 
by \citet{Craig1987}). 

It is more difficult to compare predictions of dynamic topography with 
observations. The reason for this is that there are significant long-wavelength 
features in the Earth's topography that are not associated with 
mantle convection e.g. the difference in elevation between the oceans and the 
continents, and other topography associated with variations in crustal 
thickness. A recent attempt has been made to estimate the power spectrum of 
dynamic topography by \citet{Hoggard2016}, by fitting spherical harmonics to 
point observations of residual depth in the oceans, and a fixed scaling of 
long-wavelength gravity anomalies in the continents. Their estimate of power 
indicates a Kaula-rule-like decay in the power spectral density as $k^{-2}$, 
and is plotted in \autoref{fig:grav_spec}(b). At wavelengths longer than 2000 
km 
the convection experiments are not able to explain the \citet{Hoggard2016} 
estimate of dynamic topography. At shorter wavelengths  spectra become more 
comparable, although only a limited comparison can be made because the 
\citet{Hoggard2016} estimate is limited to spherical harmonic degree 30 (a 
wavelength of 1300 km). The \citet{Hoggard2016} estimates are based on 
air-loading in the continents and water-loading in the oceans. The power 
spectra of the convection experiments in plotted in \autoref{fig:grav_spec}(b) 
assume air loading. The effect of water loading would be to increase the 
amplitude of the power spectral density of the topography by a factor of $(1- 
\rho_w/\rho_0)^2 \approx 2$, which would represent only a small shift on the 
log-scale plot of \autoref{fig:grav_spec}(b). 

In addition to comparing the observed and calculated individual spectra of 
gravity and topography, it is also of interest to look at their relationship 
to one another \citep{Parsons1983}. This relationship is typically 
characterised in terms of the admittance and coherence of the two signals, 
with the topography taken as input and the gravity as output 
(appendix \protect{\ref{sec:admittance}}). 
\autoref{fig:grav_spec}(c) shows the admittance in the spectral domain, 
calculated from the numerical experiments. The value of the admittance 
increases with increasing wave number (decreasing wavelength), approximately 
as $\log(k)$ for the range of wave numbers shown. The behaviour of admittance 
with wave number is similar for the different boundary conditions over the 
range of interest, with significant departures only noticeable at long 
wavelengths ($>$2000~km).  Indeed, at long wavelengths the admittance for the 
free-rigid case becomes negative around a wavelength of 3700 km (as noted by 
\citet{Parsons1983}, see their Figure 5). \autoref{fig:grav_spec}(c) is 
consistent with the behaviour in the cross-plots of \autoref{fig:cross_plots}. 
The gravity anomalies have most power at wavelengths around 1000~km,  
corresponding to an air-loaded admittance of around 45 mGal~km$^{-1}$ in  
\autoref{fig:grav_spec}(c), which is broadly the slope obtained from the 
cross-plots.    The corresponding coherence in \autoref{fig:grav_spec}(d) is 
close to 1, reflecting the good correlation between the two observables that 
can be seen in the space domain plots. Only at wavelengths longer than 2000~km 
is a weak coherence between the signals seen, and then only significantly for 
the free-rigid case. This behaviour mirrors the admittance at long-wavelengths 
for the free-rigid case, where gravity and topography correlate positively 
except for wavelengths longer than 4000~km when they correlate negatively.

The frequency domain plots in \autoref{fig:grav_spec} illustrate the 
spectral properties of the gravity and topography at the top of the convecting 
box, as shown in the space domain in Figures \ref{fig:grav_nofilter} and 
\ref{fig:topo_nofilter}. When considering signals at the Earth's surface, the 
spectral properties will be further modified by the filtering effect of the 
MBL on top. How significant this effect is depends on the effective elastic 
thickness and the thickness of the MBL. For an elastic thickness $T_e = 30$~km 
as used in Figures \ref{fig:grav_filter} and \ref{fig:topo_filter}, the 
wavelength at which the Fourier coefficients of topography are reduced by a 
factor of 2 is $\lambda^{1/2}_\text{flex} = 330$ km. Wavelengths shorter than 
this will be significantly attenuated by the flexural filtering; wavelengths 
longer than this will not. The topography spectra in \autoref{fig:grav_spec} 
would only be significantly different at wavelengths shorter than 
$\lambda^{1/2}_\text{flex}$ were a flexural filter to be applied. The effect 
of flexural filtering on the gravity is more complicated, and acts to produce 
a modest increase in the gravity signal in a wavelength band around that 
associated with the MBL thickness and that associated with flexure (see 
appendix \ref{sec:flexfilter} and 
\protect{\autoref{fig:grav_spec_flex_filter}} for further discussion). 

\section{Melt generation}\label{sec:melt}
The generation of melt, its separation from its source 
regions and the time $\tau$ required for it to move from its source to the 
surface,
have all been extensively studied, both theoretically using two-phase flow 
equations, and observationally, using a variety of geochemical approaches.  
The two-phase flow equations show that basaltic melt separates from its source 
regions when the melt fraction by volume $\phi_0$ exceeds $\sim 0.5$\%  
\protect{\citep{McKenzie1985}}.  Studies of the composition of abyssal 
peridotites \protect{\citep{Johnson1990,Warren2016}} show that the incompatible 
elements that were present in 
the source before melting occurred have been removed by the melt.  Estimates 
of the melt fraction present during the melting are between 0.2 and 0.7\% 
\protect{\citep{Slater2001,Liang2010}}.  Estimates of $\phi_0$ and $\tau$ 
can also be obtained from measurements of U-series disequilibria 
\protect{\citep{McKenzie1985c,Kokfelt2003,Stracke2006,Koornneef2012,Turner2016}}
.  Most estimates give $\phi_0 \leq 0.5$\% and $\tau\le 1$ ka.  Perhaps 
the strongest constraint on the values of both $\phi_0$ and $\tau$ comes from 
modeling the generation of melt by deglaciation.  When most of the ice 
covering Iceland melted at the end of the last glaciation the melt production 
rate suddenly increased \citep{Maclennan2002,Eason2015}.  The 
thickness of the ice that melted was about 2 km, increasing the melt fraction 
present in the source region by only about 0.2\% 
\protect{\citep{Jull1996,Eksinchol2019}}.  This increase was sufficient to 
generate large shield 
volcanoes within about 1 ka of the removal of the ice.  These models and 
observations all show that melt generated by decompression melting in the 
upper mantle rapidly moves to the surface, and that no appreciable volume 
remains in the source region.  We therefore calculated the rate of melt 
production by simply vertically-integrating the melting rate over the thickness 
of the 
layer in which melt was being produced.

\autoref{fig:melt_gen} shows the calculated 
rate of melt production for the four 
$\text{Ra}=10^6$ simulations, with a lithospheric 
thickness of 80 km, and assuming various water contents. The 
hydrous melting parametrisation of \protect{\citet{Katz2003}} was used 
(appendix 
\protect{\ref{sec:ptemp}}).  Melt production in \protect{\autoref{fig:melt_gen}} 
only occurs 
where the mantle is upwelling, where the gravity anomalies and 
topography are positive (Figures \protect{\ref{fig:grav_filter}} and 
\protect{\ref{fig:topo_filter}}).  However, the spatial extent of 
the regions 
of high melt rate are smaller than the regions of positive gravity anomaly. 
Most of the melting takes place in the regions directly above the narrow 
($\sim 60$ km diameter) upwelling plume conduits, where the decompression rate 
is greatest. However, melting also takes place, albeit to a lesser 
degree, in a broader region around the conduits and above some of the rising 
sheets connecting neighbouring plumes (i.e. on the spokes as well as the hubs 
of the spoke-pattern). Melt generation along some of the spokes is 
particularly clear for the free-free simulation, with linear bands of 
relatively low melt production connecting concentrated centres of relatively 
high melt production.   

\begin{figure} 
\includegraphics[width=\columnwidth]{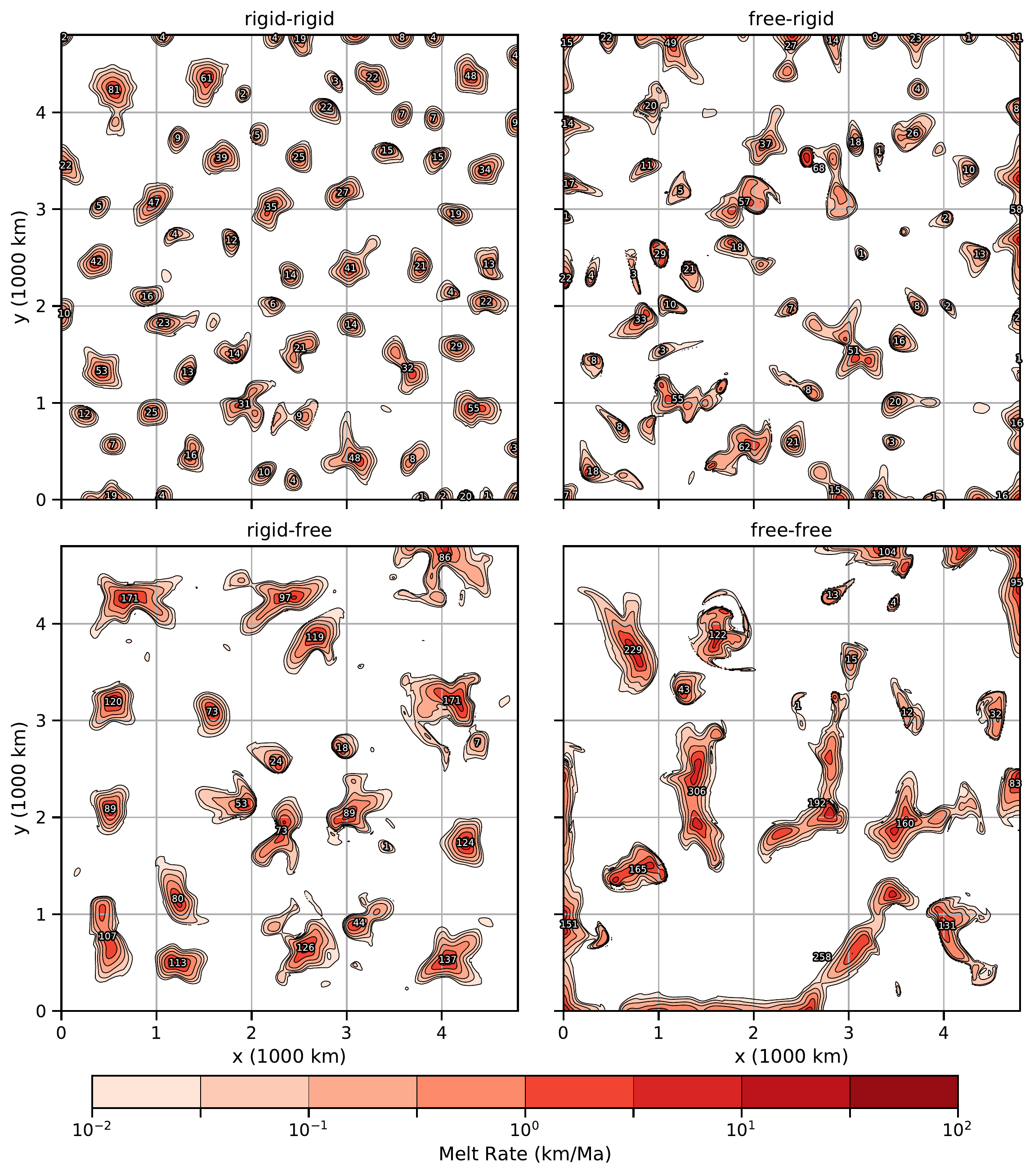}
\caption{Vertically integrated rate of melt production beneath an 80 km thick 
lithosphere with 100 ppm water. Note that the colourscale for melt rate is 
logarithmic. Numbers in white give the total rate of melt production in 
km$^3$~ka$^{-1}$ for each contiguous zone of melting. Regions of total melt 
production less than 1 km$^3$~ka$^{-1}$ are not labelled.}
\label{fig:melt_gen}
\end{figure}

\begin{figure}
 \includegraphics[width=0.9\columnwidth]{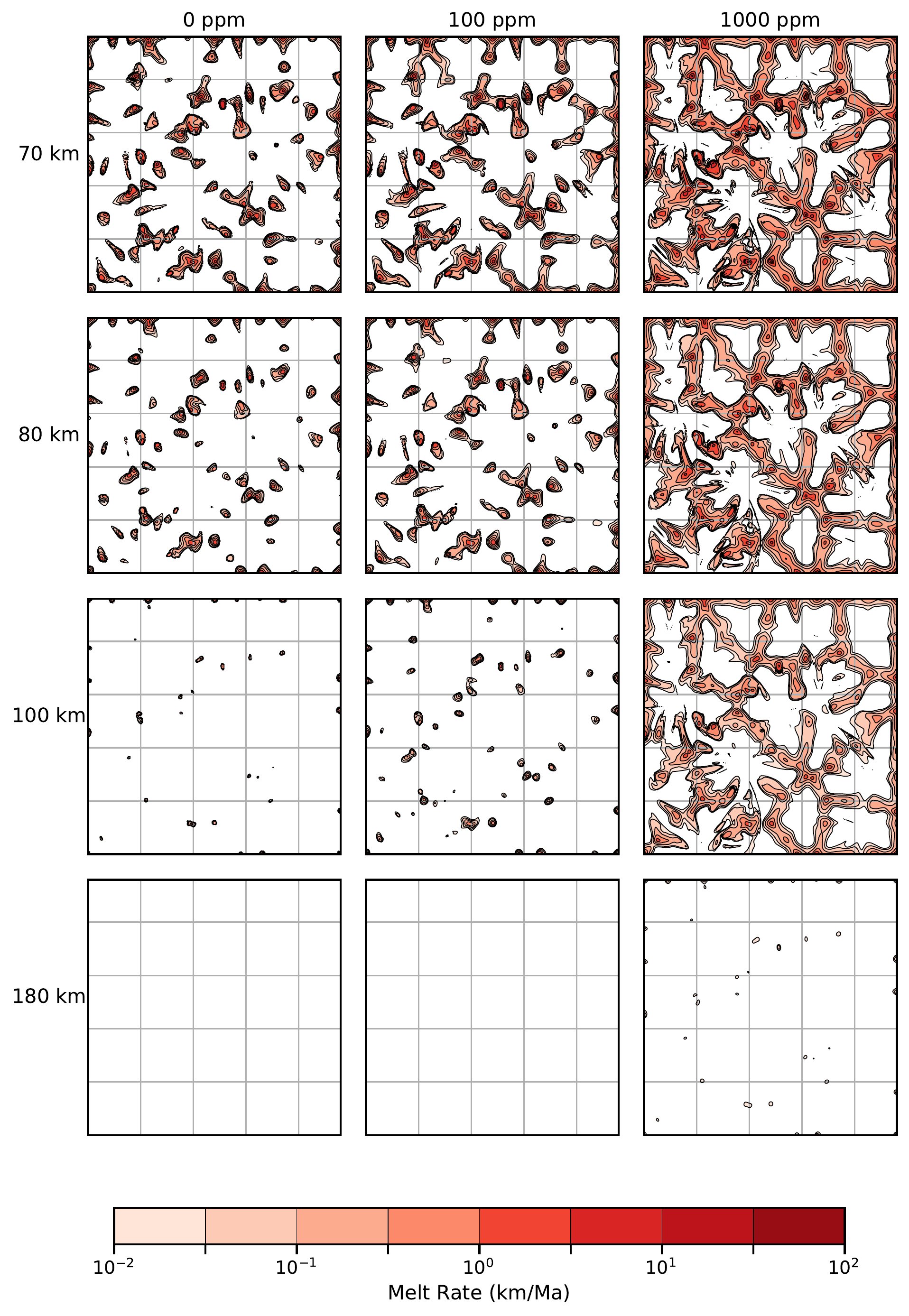}
\caption{Vertically integrated rate of melt production for the free-rigid 
simulation with $\text{Ra}=10^6$, 
showing the effect of varying the water content (as 
labelled horizontally in ppm), and different lithospheric thicknesses (as 
labelled vertically in km).}
 \label{fig:melt_gen_controls}
 \end{figure}

The rate of melt production is particularly sensitive to the thickness of the 
lithosphere and the water content of the mantle. 
\autoref{fig:melt_gen_controls} 
illustrates the effect of varying these two parameters for the free-rigid 
$\text{Ra}=10^6$ simulation. At low water contents, and beneath thick 
lithosphere, melting is restricted to the hubs in the spoke-pattern, if indeed 
melting happens at all. However, high water contents and thin 
lithosphere result in melting along the spokes. As 
\autoref{fig:melt_gen_controls} 
illustrates, for lithosphere as thick as 180 km, melting is suppressed unless 
the water content is sufficiently high ($\sim 1000$ ppm). For lithosphere as 
thin as 70 km, melting along both the hubs and the spokes can be seen even for 
modest ($\sim 100$ ppm) water contents. The water content of the convecting 
upper mantle is probably between 100 and 200 ppm 
\protect{\citep{Michael1995,Saal2002}}
. In 
contrast the water content is likely to be considerably 
greater where the base of the lithosphere has been enriched by metasomatism.  
It is not straightforward to estimate the water content of such regions using 
that in the nodules brought up by magmas such as kimberlites from depths of 
100-200 km, because they are often infiltrated by the host magma. Protons are 
especially mobile.  More reliable estimates can be obtained from the Ce 
concentration, because Ce and H have similar bulk partition coefficients 
between magma and peridotite \protect{\citep{Aubaud2004}}, and the Ce 
concentration in 
nodules is less affected by infiltration than is that of H 
(\protect{\citet{Erlank1987}} 
p 283).  The Ce concentration in the commonest class of nodules is $\sim 10$ 
ppm \protect{\citep{Erlank1987}}, or about $10 \times$ that of the convecting 
upper 
mantle.  Therefore the metasomatically enriched region at the base of thick 
old lithosphere probably has a water concentration of $\sim 1000$ ppm. Where 
the lithosphere is thin, \autoref{fig:melt_gen_controls} shows that such 
high water 
concentrations will lead to widespread melting along spokes, and at the hotter 
hubs melting at depths as great as 180 km.

\section{Terrestrial Observations}\label{sec:obs}

The numerical experiments described above show that the planform of mantle 
convection will be most obviously expressed in the surface observables when 
both the elastic thickness and lithospheric thickness are small.  
\autoref{fig:Fig_16} shows 
that the lithospheric thickness exceeds 120 km over large regions of western 
and southern Africa. In these regions the volcanism consists of small-volume 
alkalic eruptions, such as kimberlites, that contain high concentrations of 
carbonates and hydrous minerals.  Where kimberlites are diamond-bearing, 
\autoref{fig:Fig_16} shows that the lithospheric thickness generally exceeds 
150 
km.  The 
limited spatial resolution of surface wave tomography, of $\sim 250$ km, 
probably accounts for the few diamond-bearing locations in \autoref{fig:Fig_16} 
that appear 
to have thinner lithosphere.

\begin{figure}
\centering
\includegraphics[width=\columnwidth]{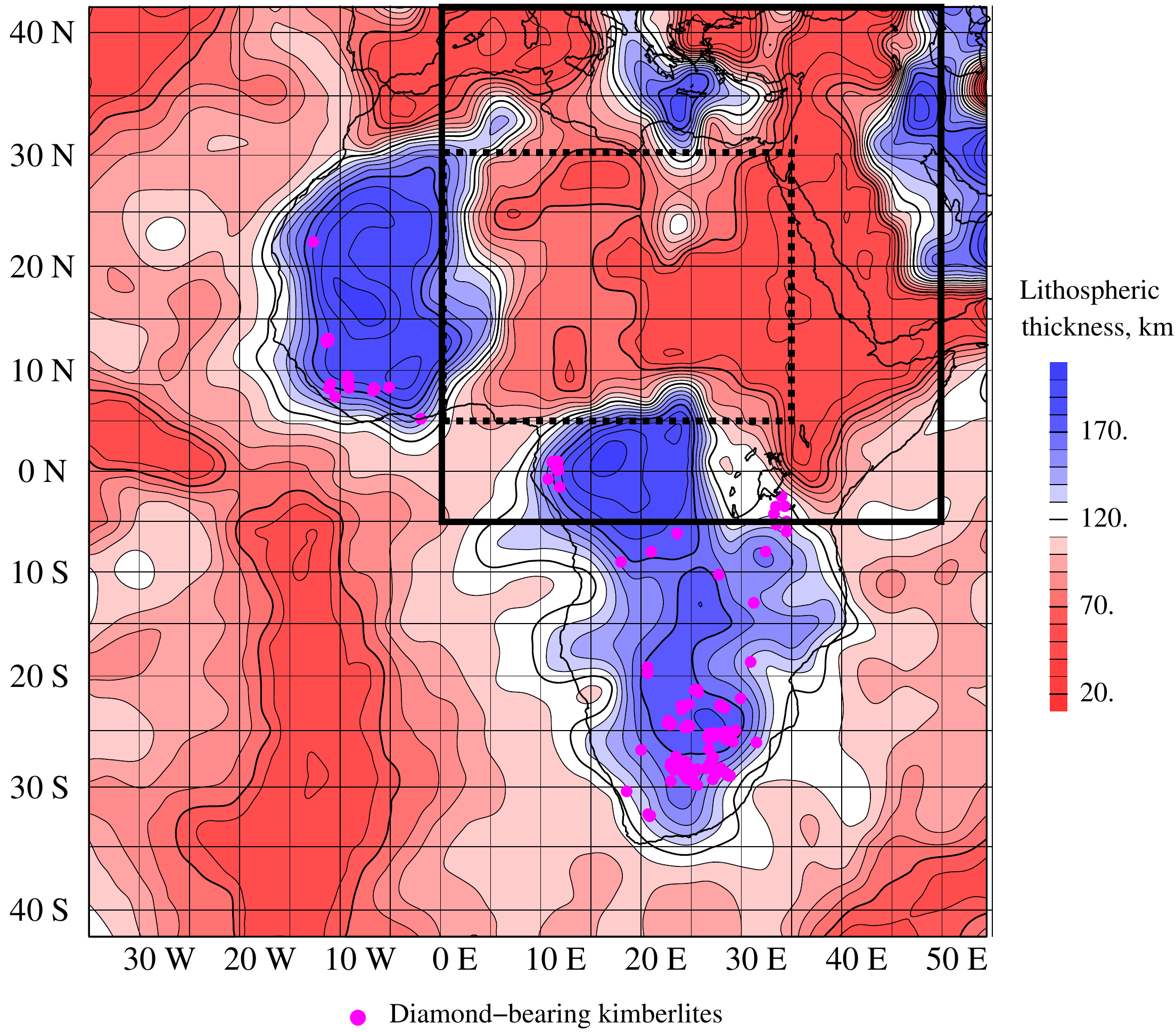}
\caption{Lithospheric thickness calculated from surface wave 
tomography \citep{Priestley2018}. The thick square box indicates a region that 
is of the same horizontal extent as the convection simulations. The elastic 
thickness of the region within the dotted lines is likely to be less than 4~km 
(\autoref{fig:Fig_17}).}
\label{fig:Fig_16}
\end{figure}

\begin{figure}
\centering
\includegraphics[width=0.88\columnwidth] {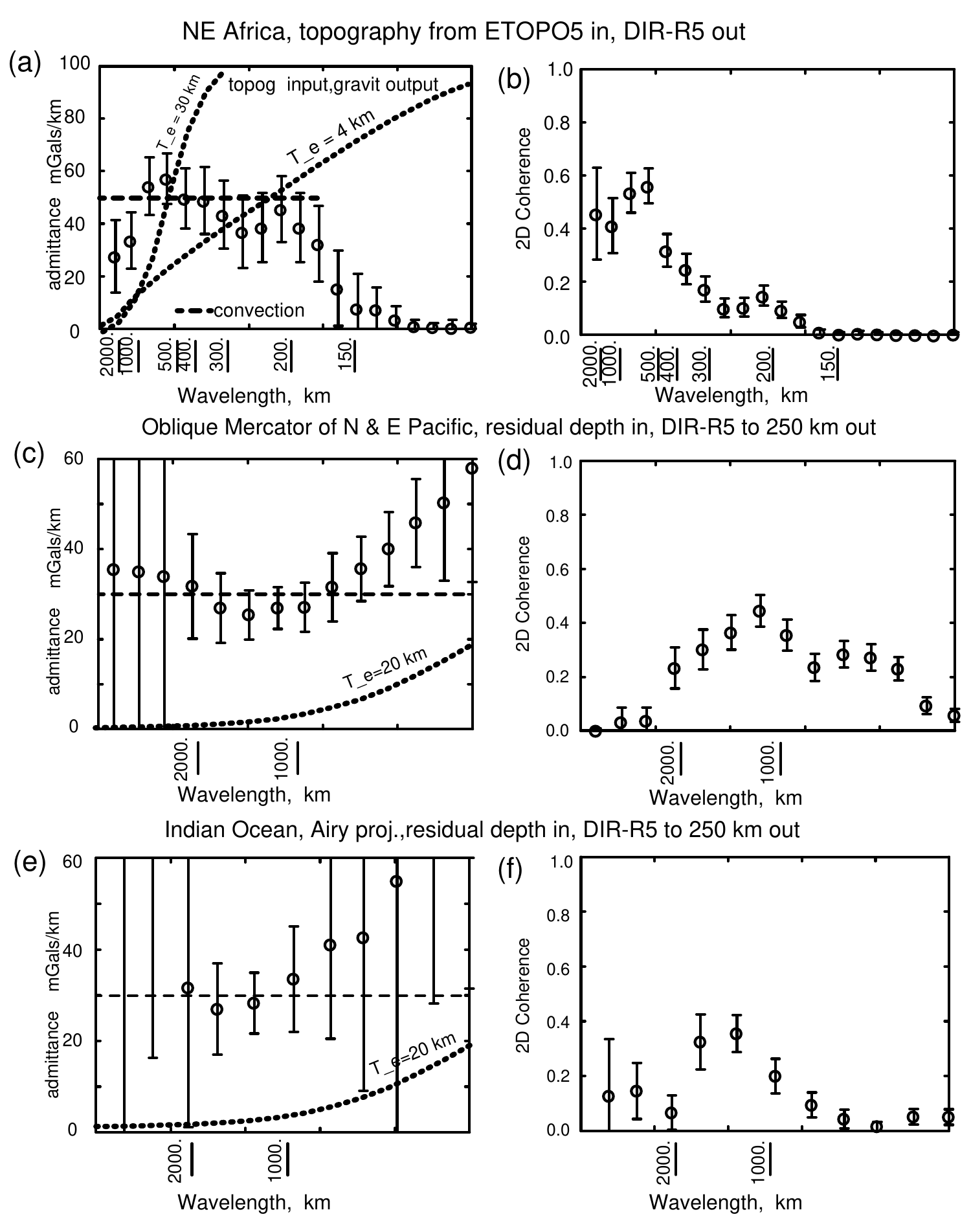}
\caption{Plots of the admittance and coherence of gravity and 
topography 
versus $1/\lambda$ where $\lambda$ is the 
wavelength and $k= 2 \pi/\lambda$ the wavenumber, in three regions.  (a) 
and (b) use free 
air gravity anomalies and 
topography from inside the box in \autoref{fig:Fig_16} marked by the dotted 
line. The gravity field was calculated from DIR-R5 \citep{Bruinsma2014} by 
setting the coefficients 
from $l=2$ to 7 to 0, and  applying a taper $f,=(l-7)/5$, to those from $l=8$ 
to 11 (wavelengths 3333 to 5000 km).  A low pass filter falling to 1/2 at 250 
km was applied to remove the short wavelength anomalies associated with 
elastic flexure.  The topography is taken from ETOPO5. The admittance was 
calculated using the topography as input, gravity as output.  The two dotted 
lines show the flexural admittance for two values of the elastic thickness 
$T_e$. (c)-(e) show corresponding plots for the Pacific (see supplementary 
material for maps) and the Indian Oceans (see \autoref{fig:Fig_19}).  The 
gravity anomalies were calculated from the DIR-R5 coefficients with those of 
degree 2 set to 0.  The admittances in (c) and (e) were calculated from the 
ratio of the spectral coefficients.  The dotted lines show the flexural 
admittance for $T_e=20$ km.
}
\label{fig:Fig_17}
\end{figure}

The lithospheric thickness within much of the rectangle marked by the thick 
continuous black line in \autoref{fig:Fig_16} is less than 80 km, and the 
horizontal extent of this region is similar to that of the numerical 
experiments.  In the eastern and northern parts of the area there are 
sufficient surface gravity measurements to allow $T_e$ to be estimated 
from the transfer function between the free air gravity 
anomalies and the topography \protect{\citep{McKenzie1997}}, giving 
values of 3-4 km (see supplementary material). The coherence method and 
Bouguer anomalies provides only an upper bound, not an estimate, of the value  
of $T_e$ \protect{\citep{McKenzie2016}}.  Within the box marked 
by the dotted lines in \autoref{fig:Fig_16} there are few surface gravity 
measurements.  Instead the satellite gravity field DIR-R5 can be used to show 
that the admittance is about $50\ {\rm mGal\ km}^{-1}$ between wavelengths of 
200 and 1000 km, and that the elastic thickness is probably less than 4 km 
(\autoref{fig:Fig_17}(a) and (b)). Therefore in this region and wavelength 
band both the gravity and topography are controlled by convection.      
The same is not the case in southern Africa, where the elastic thickness is 
about 30 km \citep{McKenzie2015}. 

\autoref{fig:Fig_18} shows maps of gravity, topography, and 
subaerial volcanism across 
Africa and Arabia. The volcanism beneath the Red Sea and in 
the Afar results from upwelling of the mantle between separating plates. 
However, in Ethiopia, Kenya and Kivu the upwelling from the limited extension 
is insufficient to account for the extensive volcanism. There is no obvious 
orientation of the topography and gravity anomalies, probably because Africa 
is almost stationary with respect to the hotspot frame.
The correspondence between the features in 
\autoref{fig:Fig_18} and the maps 
from the numerical experiments is striking.  Volcanism is almost entirely 
restricted to regions where the lithospheric thickness is less than 70 km.  
The only clear exception is Kivu, where the spatial resolution of the surface 
wave tomography is probably insufficient to resolve the thickness of the thin 
lithosphere beneath the Western Rift.  The linear volcanic feature extending 
from Kenya to the Kars Plateau resembles similar linear features in 
\autoref{fig:melt_gen}, with localised regions of concentrated upwelling being 
associated with positive gravity and elevation, and with enhanced volcanism.  
Like the numerical experiments, the volcanism is more localised than are the 
associated positive gravity and topographic anomalies. What is less clear is 
which of the four combinations of boundary conditions best fits the 
observations.  Figures \ref{fig:grav_filter} and \ref{fig:topo_filter} show 
that the observed horizontal scales of the 
convective features are probably larger than those of the rigid-rigid, and 
smaller than those of the free-free, experiments.  The scales of the anomalies 
in the other two experiments are similar, both to each other and to the 
observed scales. The rigid-free experiment has broadly circular patterns of 
positive gravity anomalies surrounded by linear negative anomalies, whereas 
the free-rigid case has the opposite.  The 
viscosity of the lower mantle is greater, and that of the asthenosphere 
immediately below the lithosphere less, 
than that of the upper mantle.  These viscosity variations suggest that the 
free-rigid experiment should match the observed patterns better than the 
rigid-free case.  However, the patterns in \autoref{fig:Fig_18} are not 
obviously more like the free-rigid features than the rigid-free ones.  
Furthermore the variation of viscosity with temperature, which has been 
ignored, may have a strong influence on the geometry, and in particular 
whether the planform is dominated by rising or sinking fluid in the hubs. At low 
Rayleigh numbers the answer to this question is 
controlled by the sign of $\d\eta/dT$, with the flow in the plumes being in 
the direction of increasing viscosity \citep{Segel1962}. 
If the same is 
true at large Rayleigh numbers the hubs in both these experiments will consist 
of hot rising material.

The Cameroon Line forms 
a curve, similar to features in \autoref{fig:melt_gen}.  As the 
lithosphere thickness increases to the SW, where the Line lies beneath 
Atlantic lithosphere, the volcanism decreases.  The association of positive 
gravity anomalies, elevated topography and volcanism is clearly expressed even 
in relatively small features like A\"{i}r and Darfur.  Many volcanic centres,
such as A\"{i}r and Hoggar, and Haruj and Tibesti, are linked to each 
other by lines of positive gravity and topography, where 
there is limited volcanism.  An especially obvious feature 
extends from S. Arabia to Anatolia, where the volcanism is beneath 
Western Arabia, not the Red Sea.  Such lines are most clearly visible in 
\autoref{fig:Fig_18}(c) and (e) where the lithosphere is thin beneath NE 
Africa, Arabia and Anatolia.  All these features are similar to those of the 
numerical experiments, and all are consistent with a spoke pattern of 
convection existing beneath the region.  In particular, and as expected from 
\autoref{fig:melt_gen}, the extent of the volcanism is controlled by 
variations in lithospheric thickness, and is limited in the south and east by 
the thick lithosphere of the Congo Craton.

The Cameroon Line and the line of active volcanism that extends from Afar to 
eastern Anatolia have long puzzled geophysicists, because their geometry is 
difficult to reconcile with a planform of mantle convection consisting of 
plumes. \citet{Sleep1997,Sleep2008} and \citet{Ebinger1998} argued that these 
linear volcanic features were produced by lateral flow from plumes in channels 
beneath the lithosphere.  One problem with this proposal is that the 
composition of the volcanics along the Cameroon Line shows so little variation 
\citep{Fitton1987,Lee1994}.

A different model was proposed by \citet{Milelli2012} which emphasised the 
location of the volcanism, which has remained in the same region of Africa as 
the continent has moved.  They argued that this behaviour required the 
volcanism to result from thermal instabilities in the lower part of the 
lithosphere, rather than being the surface expression of convective upwellings 
in the upper mantle below the plates. The numerical experiments described 
above show that the expected planform of upper mantle convection is that 
of hubs joined by spokes, both of which can generate melt if the lithosphere 
is sufficiently thin.  The observed linearity of the Cameroon Line and other 
features in NE Africa therefore requires no special explanation.  The 
experiments also show that \citet{Milelli2012}'s observations may also have a 
simple 
explanation, since the melting rate, and not the planform, is controlled by 
the lithospheric thickness, and the volcanism of the Cameroon Line lies along 
the northern edge of the thick lithosphere of the Congo Craton.  In contrast 
the line of volcanism from the Afar to Kars lies within a region of relatively 
uniform lithospheric thickness.  It is therefore unlikely that such linear 
features all form from edge convection like that discussed by 
\citet{King1998}.

\begin{figure}
\includegraphics[width=\columnwidth] {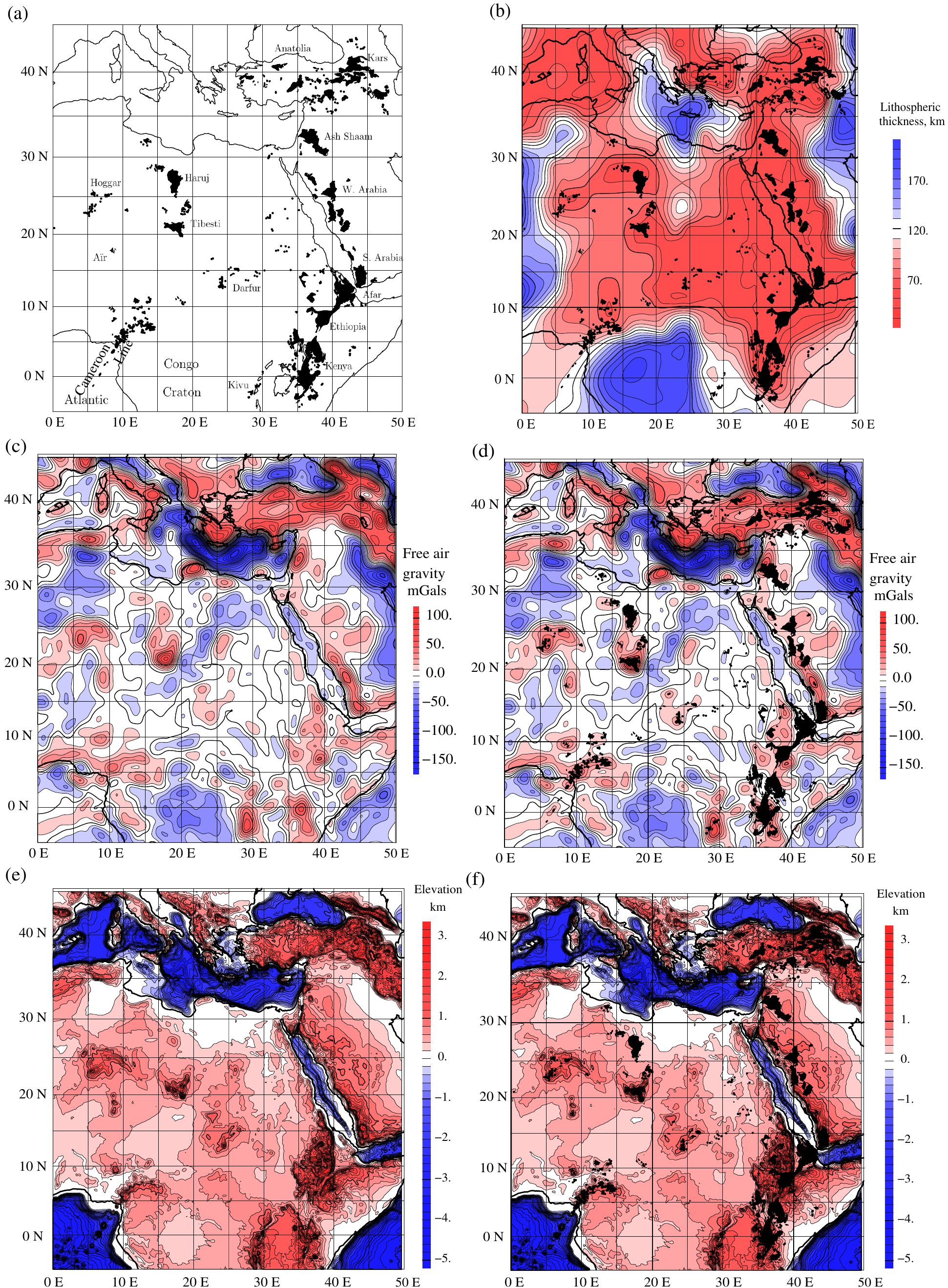}
\caption{Maps of the volcanism, lithospheric thickness, gravity, and 
topography within the region marked by the heavy black line in 
\autoref{fig:Fig_16}.    The gravity and topography are shown without 
(\autoref{fig:Fig_18}(c) and (e)) and with (\autoref{fig:Fig_18}(d) and (e)) 
the regions covered by volcanics, most of which are Miocene or younger, taken 
from \citet{Thorpe1974}, \citet{Ball2019}, and \url{earthwise.bgs.ac.uk}.
} 
\label{fig:Fig_18} 
\end{figure} 

In general it is not possible to compare the melt generation rates in 
\autoref{fig:melt_gen} with those observed because they are so rarely 
estimated by the geologists who map the volcanics.  An exception is Mount 
Cameroon, which is the most active volcano in Africa. Its eruption rate was 
estimated by \citet{Suh2003} to be about 
$700\ {\rm km}^3\ {\rm /Ma}$.  When the lithospheric thickness is 80 km the 
larger hubs in the free-rigid experiment in \autoref{fig:melt_gen} produce 
about 
$6\times 10^4\ {\rm km}^3\ {\rm /Ma}$ and the smaller ones 
$1\times 10^3\ {\rm km}^3\ {\rm /Ma}$.  The rates of melt generation in the 
numerical experiments can therefore easily account for the observed rates.  
But they are quite inadequate to account for the production rates that occur 
during major flood volcanism, which commonly exceed 
$1\times 10^6\ {\rm km}^3\ {\rm /Ma}$. Like the long wavelength gravity 
anomalies discussed below, simple isoviscous upper mantle convective models 
cannot account for such events.

The box in \autoref{fig:Fig_16} is too small to be used to study the long 
wavelength components (wavelengths $>$ 1000 km)  of the Earth's dynamic 
topography and gravity.  These are best studied in the Indian and Pacific 
Oceans.  The elastic thickness of old oceanic lithosphere is about 20 km (e.g. 
\citet{McKenzie2014}).  Therefore wavelengths greater than 
about 800 km are 
little affected by the thickness and elastic properties of the lithosphere 
(\autoref{fig:Fig_17}(c) and (e)).  In many continental regions the surface 
topography is dominated by isostatically compensated variations in crustal 
thickness.  Such structures are less common in oceanic regions, where the 
bathymetry is principally controlled by the age of the lithosphere.  The 
effect of plate cooling can be removed by using a depth-age model, and regions 
of thick crust removed by hand 
\citep[e.g.][]{Crosby2006,Crosby2009,Hoggard2017}.  The resulting residual 
topography 
is then largely 
supported by convection.  \autoref{fig:Fig_17} shows a comparison between free 
air gravity, calculated from DIR-R5 with the coefficients of degree 2 set to 
zero, and Crosby's values of residual depth in the Indian and Pacific Oceans.  
Those obtained from Hoggard's values of residual depth are similar 
and are 
shown in the supplementary material.
\autoref{fig:Fig_19} shows maps of such anomalies in the Indian Ocean: those 
for the Pacific are illustrated in the supplementary material.  The admittance 
between wavelengths of 1000 and 2000 km is about 30 mGal/km, in agreement with 
the values from the numerical experiments in 
\protect{\autoref{fig:grav_spec}}. 
However, at wavelengths greater than about 2000 km the gravity and residual 
depth anomalies in both oceans cease to be coherent.  This incoherency is 
particularly striking in the Indian Ocean, where the large negative gravity 
anomaly covering the NE part of the Ocean (\autoref{fig:Fig_19}(a)) has no 
expression in the residual depth (\autoref{fig:Fig_19}(b)). At wavelengths 
greater than 2000 km the observed power spectrum also differs from that 
calculated from the box models (\autoref{fig:grav_spec}(a)). The power in the 
observed gravity field continues to increase at wavelengths longer than 2000 
km, unlike that from the numerical experiments.  This behaviour shows that 
simple isoviscous convective models cannot account for the longest wavelength 
part of the Earth's gravity field, and suggest that it is not maintained by 
upper mantle convection. Though the size of the boxes used for the numerical 
experiments is too small to determine the power at such 
wavelengths accurately, there is no suggestion in the planforms that the 
presence of lateral boundaries governs the scale of the convection.  

\begin{figure}
\centering
\includegraphics[width=0.81\columnwidth]{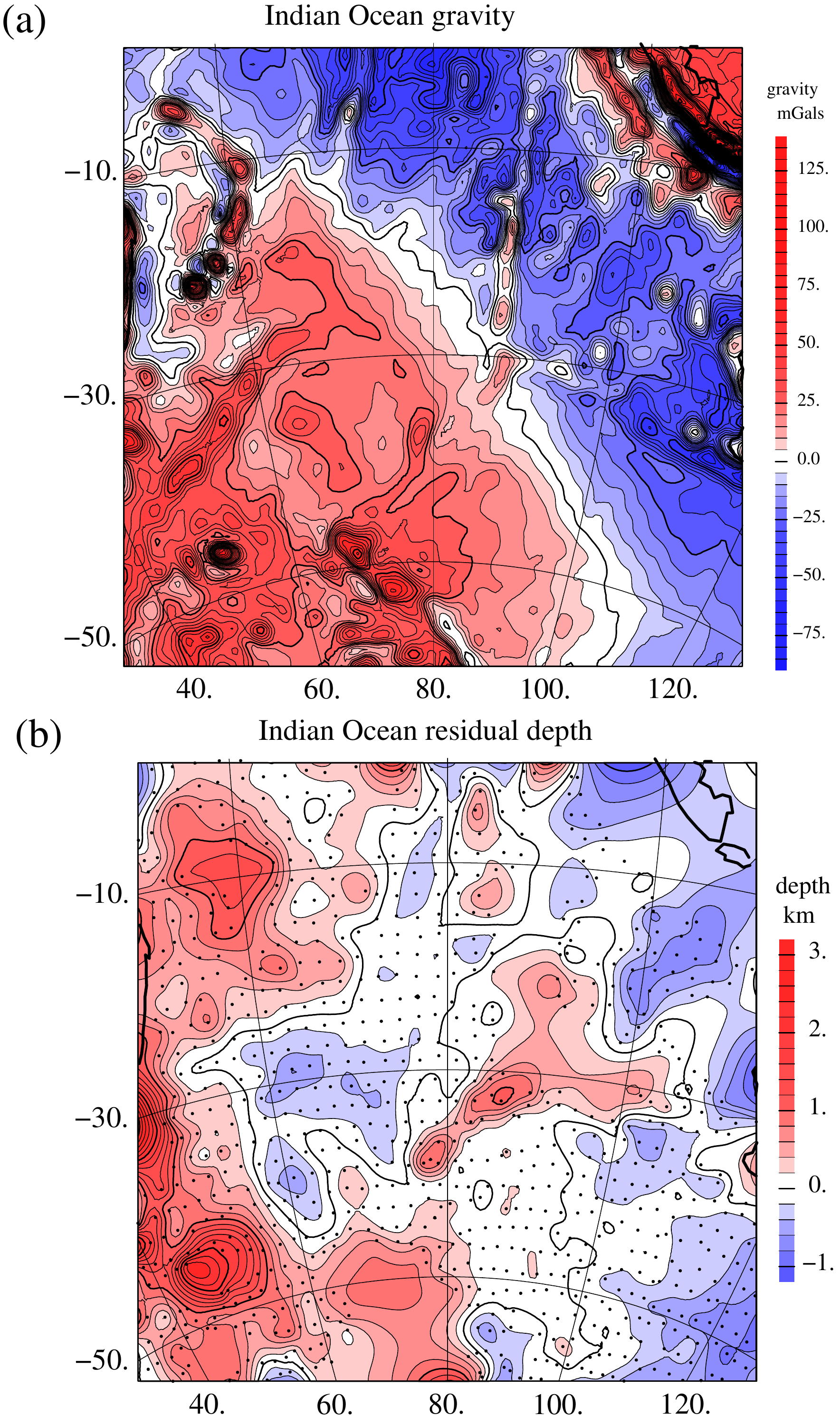}
\caption{Gravity and residual depth in the Indian Ocean. (a) Gravity 
from DIR-R5, with coefficients 
$l=2$ set to 0 and a filter applied, falling to 1/2 at 250 km, to remove the 
short wavelength components.  (b) Residual 
depths, averaged over $2^\circ \times 2^\circ$ boxes \citep{Crosby2006}. The 
dots show the locations of the resulting averages.   
Airy projection with centre $-30^\circ$N, $80^\circ$E. $\beta=30^\circ$.
}
\label{fig:Fig_19}
\end{figure}

\section{Conclusions}\label{sec:conclude}

The numerical experiments described above show that the observed gravity and 
topographic anomalies are reproduced by the simplest isoviscous fluid 
dynamical model of thermal convection. The wavelength at which convective 
support dominates elastic support is controlled by the elastic thickness 
$T_e$, and varies from about 200 km where $T_e \leq 4$ km in NE Africa to 
$\sim 500$ km in southern Africa, where $T_e\sim 30$ km.  Melt generation 
occurs where mantle material moves upwards, and is therefore controlled by the 
lithospheric thickness, and not by the value of $T_e$.  The correspondence 
between the volcanism and the gravitational and topographic anomalies in NE 
Africa is striking, and shows that they all result from the convective 
circulation. 

There is no similar correspondence between the results from the numerical 
experiments and the gravity and residual depth anomalies at wavelengths 
greater than 2000 km. Furthermore the absence of correlation between residual 
depth and gravity anomalies with wavelengths greater than 2000 km in the 
Pacific and Indian Oceans is unlike the behaviour observed at shorter 
wavelengths in these oceans.  The simple numerical models discussed here cannot 
account for 
the long wavelength gravity anomalies with spherical harmonic degrees $l \leq 
20$. 

The close correspondence between the calculated and observed topography, 
gravity and volcanism suggests that it should be possible to use the surface 
observables where the lithosphere is thin and $T_e$ is small, together with 
the isoviscous convective equations, to map the convective circulation in the 
upper mantle.

\section*{Appendix}
\appendix

\section{Numerical methods}

\subsection{Gravity, topography, 
and flexure}\label{sec:flexfilter}

To calculate surface gravity and topography, the second-order finite 
element temperature 
fields were first linearly interpolated onto a regular grid of points with 64 
mesh 
points in the vertical and 512 in the horizontal. The method of 
\citet{Parsons1983} and \citet{Craig1987} was then used to calculate both 
topography and gravity, by 
Fourier transforming the grids of data, multiplying by an appropriate 
filter (given in Appendix A of \citet{Parsons1983} for the different boundary 
conditions), and transforming back. For gravity calculations both the top 
boundary and the bottom boundary were assumed to be deformable. The 
deformation of the top and bottom boundaries was calculated using the 
approximation $(\rho_0 - \rho_w) g h = -\sigma_{zz}$, where $h$ is the 
deformation of the interface, $\sigma_{zz}$ is the normal stress at 
the relevant boundary in the fixed geometry of the convection simulations, 
and $\rho_w$ 
is the density of the fluid on the other side of the boundary.

The effect of an elastic layer above the convecting fluid is to low-pass 
filter the dynamic topography 
(\protect{\autoref{fig:grav_spec_flex_filter}b)}. To 
produce the flexurally-filtered topography 
seen in \protect{\autoref{fig:topo_filter}}, the Fourier coefficients of the 
topography 
in \protect{\autoref{fig:topo_nofilter}} were multiplied by the Fourier-domain 
flexural 
filter,
\begin{linenomath}
\begin{equation}
 F(k) = \frac{1}{1+ (\alpha_\text{flex} k)^4}, \label{eq:flexfilter}
\end{equation}
\end{linenomath}
where $k$ is the wave number, and the flexural parameter $\alpha_\text{flex}$ 
is related to the elastic 
thickness $T_e$ by
\begin{linenomath}
\begin{equation}
 \alpha_\text{flex} = \left(\frac{E T_e^3}{12 \left(1- \nu^2 \right) 
\left(\rho_0 -\rho_w\right) g} \right)^{1/4}.
\end{equation}
\end{linenomath}
The Fourier coefficients are reduced by a factor of 2 at a characteristic 
wavelength $\lambda_\text{flex} = 2 \pi \alpha_\text{flex}$. Values 
of the Young's modulus
$E=10^{11}$ Pa, and Poisson's ratio $\nu=0.25$ are assumed, such that 
$\lambda^{1/2}_\text{flex} = 330$ km for an elastic thickness $T_e = 30$ km; 
and $\lambda^{1/2}_\text{flex} = 85$ km for an elastic thickness $T_e = 5$ km. 
All plots assume air-loading and thus set $\rho_w = 0$. For water-loading 
$\rho_w = 1000$~kg~m$^{-3}$.

\begin{center}
\begin{figure}
\centering
\includegraphics[width=\columnwidth]{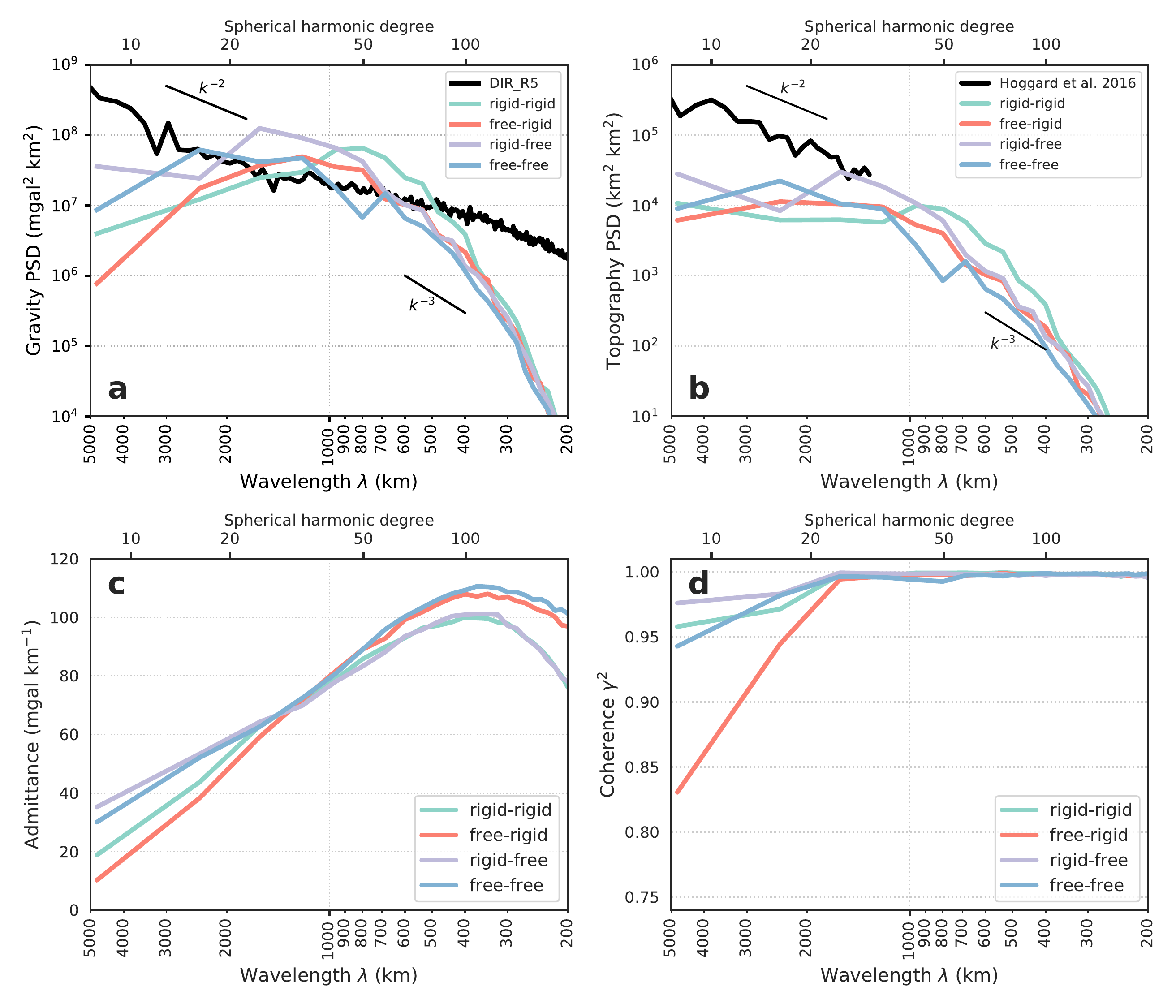}
\caption{Spectral 
characteristics as in \autoref{fig:grav_spec}, but at the surface after 
filtering through the MBL 
and an elastic layer with thickness $T_e=30$ km. Notice the significant 
decrease in the power of gravity anomalies at short wavelengths and a 
moderate increase in the power 
at intermediate wavelengths.}
\label{fig:grav_spec_flex_filter}
\end{figure}
\end{center}

The effect of an elastic layer on gravity anomalies is subtle, because 
there is a contribution to gravity anomalies both from the topography and from 
the density variations at depth, which will be attenuated by a factor $\e^{-k 
t_m}$ if there is a MBL of thickness $t_m$ on top. The gravity anomalies in 
\autoref{fig:grav_filter} are shown after this filtering process, which is 
calculated assuming the MBL is laterally uniform. The corresponding spectra are 
shown in \autoref{fig:grav_spec_flex_filter}a. If $\Delta g_0$ and $h_0$ 
represent the gravity and topography estimated at the top of the convecting 
box, 
the corresponding gravity $\Delta g_1$ at the surface can be calculated in the 
Fourier-domain from
\begin{linenomath}
\begin{equation}
 \Delta g_1(k) = \left(\Delta g_0(k) - 2 \pi G \rho_0 h_0(k) \right) \e^{-k 
t_m} + 
2 \pi G \rho_0 h_0(k)  F(k),
\end{equation}
\end{linenomath}
assuming air-loading. This equation splits the gravity anomaly into two 
components: The term on the far right represents the gravity due to the surface 
topography, which is attenuated according to the factor $F(k)$ in 
\eqref{eq:flexfilter}. The other term 
represents the density variations at depth, which are attenuated as $\e^{-k 
t_m}$ 
due to the finite thickness of the MBL. At long wavelengths gravity 
anomalies are unchanged by this additional filtering. At wavelengths 
significantly shorter than both the flexural wavelength $\lambda_\text{flex}$ 
and the wavelength $\lambda_m 
= 2 \pi t_m$ associated with the MBL,  
gravity anomalies are strongly attenuated, as both the $\e^{-k t_m}$ and 
$F(k)$ 
filters tend to zero. At intermediate wavelengths, particularly in the 
wavelength band around $\lambda_m$ and 
$\lambda^{1/2}_\text{flex}$, 
gravity anomalies actually slightly increase in magnitude due to this 
additional filtering. This behaviour occurs because a MBL acts to separate mass 
excesses at 
the surface (associated with positive gravity anomalies) from mass deficits at 
depth (associated with negative gravity anomalies). The thicker the MBL, the 
greater the attenuation of the negative anomalies, and the larger the net 
positive anomaly. Correspondingly, there is a modest increase in 
the admittance due to the addition of a MBL 
(\protect{\autoref{fig:grav_spec_flex_filter}c}; see section 2.3 of 
\citet{Crosby2009} for further discussion). It should be noted that having a 
laterally uniform thermal structure in the MBL is a poor approximation, as it 
implies a discontinuity in heat flux between the MBL and the top of the 
convecting box (where the heat flux varies laterally). However, we have made 
such an approximation here because the convection simulations fix the 
temperature at the top of the convecting layer, rather than at the Earth's 
surface. A better approach would model the temperature structure of the MBL 
during the convection simulations. However, such a modification is unlikely to 
make more than minor changes to the results of the calculations, because the 
thickness of the MBL is small compared to that of the convecting layer 
(\autoref{tab:params2}), and the temperature on the upper boundary of the MBL 
is 
fixed.  If the temperature structure of the MBL is included in the calculations 
it 
is then no longer accurate to obtain the topography and gravity for variable 
thicknesses of the MBL simply by filtering  in the spectral domain. 

\subsection{Power spectral density}\label{sec:PSD}

Power spectral density (PSD) estimates were calculated using the method 
described by 
\citet{Rexer2015}. The initial data (either gravity or topography) is a 
regularly spaced grid of points representing a region of dimensional extent 
$L_x$ 
by $L_y$. Let the number of grid points in the $x$-direction be $N$, and the 
number in the $y$-direction be $M$. The initial data is given as a matrix of 
values $d_{rs}$ where $r=0, 1, \ldots N-1$, and $s=0, 1, 
\ldots M-1$.  The convection simulations have reflecting boundary 
conditions at the sides, so the natural Fourier representation to use is a 
Discrete Cosine Transform of the first type (DCT-I), which is 
equivalent to a discrete Fourier Transform of a $2N-2$ by $2M-2$ extended grid 
of data exploiting the even symmetry. The discrete Fourier coefficients 
$f_{pq}$ are defined by equation (8) of \citet{Rexer2015},
\begin{linenomath}
\begin{equation}
 f_{pq} = \frac{1}{\left(2 N - 2 \right)\left(2 M - 2 \right) }\sum_{r = 0}^{2 
N - 2}\;  \sum_{s = 0}^{2 M - 2} d_{rs} \exp \left(- \pi i \left( \frac{s 
p}{N-1} + \frac{ r q}{M-1} \right) \right).
\end{equation}
\end{linenomath}
The even symmetry extends the data such that for $N \leq r \leq 2 N -2$ the 
value is taken from the original grid at $r^\prime = 2 N -2 -r$, and for 
$M \leq s\leq 2M - 2$ the value is taken from original grid at 
$s^\prime = 2 M -2 -s$. In dimensional variables, the corresponding wavenumbers 
of the transform are $k^x_p = \pi p/L_x$ and $k^y_q = \pi q/L_y$. Owing to 
the reflection boundary conditions, the discrete Fourier coefficients $f_{pq}$ 
are real and even in both directions. A 2D grid of power spectral density can 
then be computed from equation (10) of \citet{Rexer2015},
\begin{linenomath}
\begin{equation}
 \phi_{pq} = 4 L_x L_y | f_{pq} |^2.
\end{equation}
\end{linenomath}
Finally, the 2D-PSD were then azimuthally averaged in wavenumber space with a 
bin-width of $1.3 \times 10^{-3}$~km$^{-1}$, to produce the 1D-PSD profiles 
that 
are shown in Figures \ref{fig:grav_spec}(a) and \autoref{fig:grav_spec}(b) 
(with units of 
either mgal$^2$~km$^2$ or km$^2$~km$^2$).

Data for the Earth is typically given in terms of spherical harmonic 
coefficients, 
which need to be manipulated before they can be compared directly with the 
1D-PSD 
profiles calculated for the Cartesian geometry of the convection simulations. 
This process is also described 
by \citet{Rexer2015}. The spherical harmonic degree $l$ can be 
related to the Cartesian wavenumber $k$ by the Jeans relation approximation
\begin{linenomath}
\begin{equation}
 k = \frac{l + \tfrac{1}{2}}{R}
\end{equation}
\end{linenomath}
where $R$ is the radius of the Earth. An estimate of the 1D Cartesian PSD can 
be 
obtained from 
\begin{linenomath}
\begin{equation}
 \phi_\text{PSD} (k) = 4 \pi R^2 \frac{P_l}{2l +1}
\end{equation}
\end{linenomath}
where $P_l$ is the power at spherical harmonic degree $l$ (see 
equation (13) of \citet{Rexer2015}). 

\subsection{Admittance and coherence}\label{sec:admittance}

The admittance in \autoref{fig:grav_spec}(c) was computed as
\begin{linenomath}
\begin{equation}
 Z(k) = \frac{\left< \Delta g, h \right>}{\left< h, h \right>}
\end{equation}
\end{linenomath}
where $\left< \cdot, \cdot \right>$ represents the cross-power of the signals 
as 
a function of wavenumber $k$, calculated in the same way as the power spectra 
by 
multiplying the Fourier coefficients and then azimuthally averaging. Since the 
Fourier coefficients are real, the 
admittance is also real. The coherence in \autoref{fig:grav_spec}(d) was 
calculated 
similarly as
\begin{linenomath}
\begin{equation}
 \gamma^2(k) = \frac{\left< \Delta g, h \right>^2}{\left< \Delta g, \Delta g 
\right> \left< h, h \right>}.
\end{equation}
\end{linenomath}

\subsection{Rayleigh number scalings} \label{sec:Ra_scaling} 

\begin{table}
\begin{center}
 \begin{tabular}{ccccc}
\hline  
BC & Nu & $h^\prime_\text{RMS}$ & $\Delta g^\prime_\text{RMS}$ \\
\hline \rule{0pt}{1.0\normalbaselineskip}
rigid-rigid & $0.189\,\text{Ra}^{0.281}$ & $1.682\,\text{Ra}^{-0.342}$ &
$0.405\,\text{Ra}^{-0.308}$ \\
free-rigid &  $0.228\,\text{Ra}^{0.283}$ & $1.098\,\text{Ra}^{-0.314}$ &
$0.134\,\text{Ra}^{-0.242}$ \\
rigid-free&  $0.247\, \text{Ra}^{0.277}$ &  $1.626\,\text{Ra}^{-0.317}$ &
$0.655\,\text{Ra}^{-0.328}$ \\
free-free&  $0.253\,\text{Ra}^{0.306}$ & $1.191\,\text{Ra}^{-0.289}$ &
$0.243\,\text{Ra}^{-0.259}$\\
\hline
\end{tabular}
\end{center}
\caption{Approximate scalings with Rayleigh number for Nu, the Nusselt number 
(ratio of convective to conductive heat transfer); 
$h^\prime_\text{RMS} = h_{\text{RMS}} / \left(  \rho_0 \alpha \Delta T_p d 
/(\rho_0 -\rho_w)  \right)$, dimensionless root-mean-square dynamic topography; 
and $\Delta g^\prime_\text{RMS} = \Delta g_{\text{RMS}} / \left( 2 \pi G \rho_0 
\alpha \Delta T_p d \right)$, dimensionless root-mean-square gravity anomaly.}
\label{tab:ra_scaling}
\end{table}

Boundary layer theory suggests that there should be systematic power-law 
scalings with Rayleigh number for properties of the convecting system, such as 
the thickness of boundary layers, and the heat flux. \autoref{tab:ra_scaling} 
shows such approximate scaling laws that have been obtained from the 12 
numerical runs presented here. These scaling laws should be used with caution: 
they were determined from the properties of the system at a single snapshot in 
time, for a limited range of Rayleigh numbers. However, they illustrate the 
broad trends. As expected, Nusselt number increases with Rayleigh number: the 
behaviour for the free-free system as $\text{Nu} \propto \text{Ra}^{0.3}$ is 
close to to the $1/3$ power law expected from boundary layer theory 
\citep{McKenzie1974}. Since the dimensional scaling used in the main text 
essentially fixes the heat flux, the potential temperature difference across 
the 
layer given in \autoref{tab:params2} scales as the inverse as the Nusselt 
number scaling, i.e. approximately as $\text{Ra}^{-0.3}$. 
\autoref{tab:ra_scaling} also shows scalings for dimensionless 
topography and dimensionless gravity, which all scale roughly as 
$\text{Ra}^{-0.3}$ although there are some differences in detail. This scaling 
can be understood in broad terms from the expectation that the dynamic 
topography should be proportional to the boundary layer thickness 
\citep{Parsons1983}. Since the scaling used in \autoref{tab:ra_scaling} 
includes a $\Delta T_p$ factor, the scalings of the dimensional RMS gravity and 
topography in \autoref{tab:params2} go roughly as $\text{Ra}^{-0.6}$.

\begin{figure}
\centering
\includegraphics[width=0.7\columnwidth]
{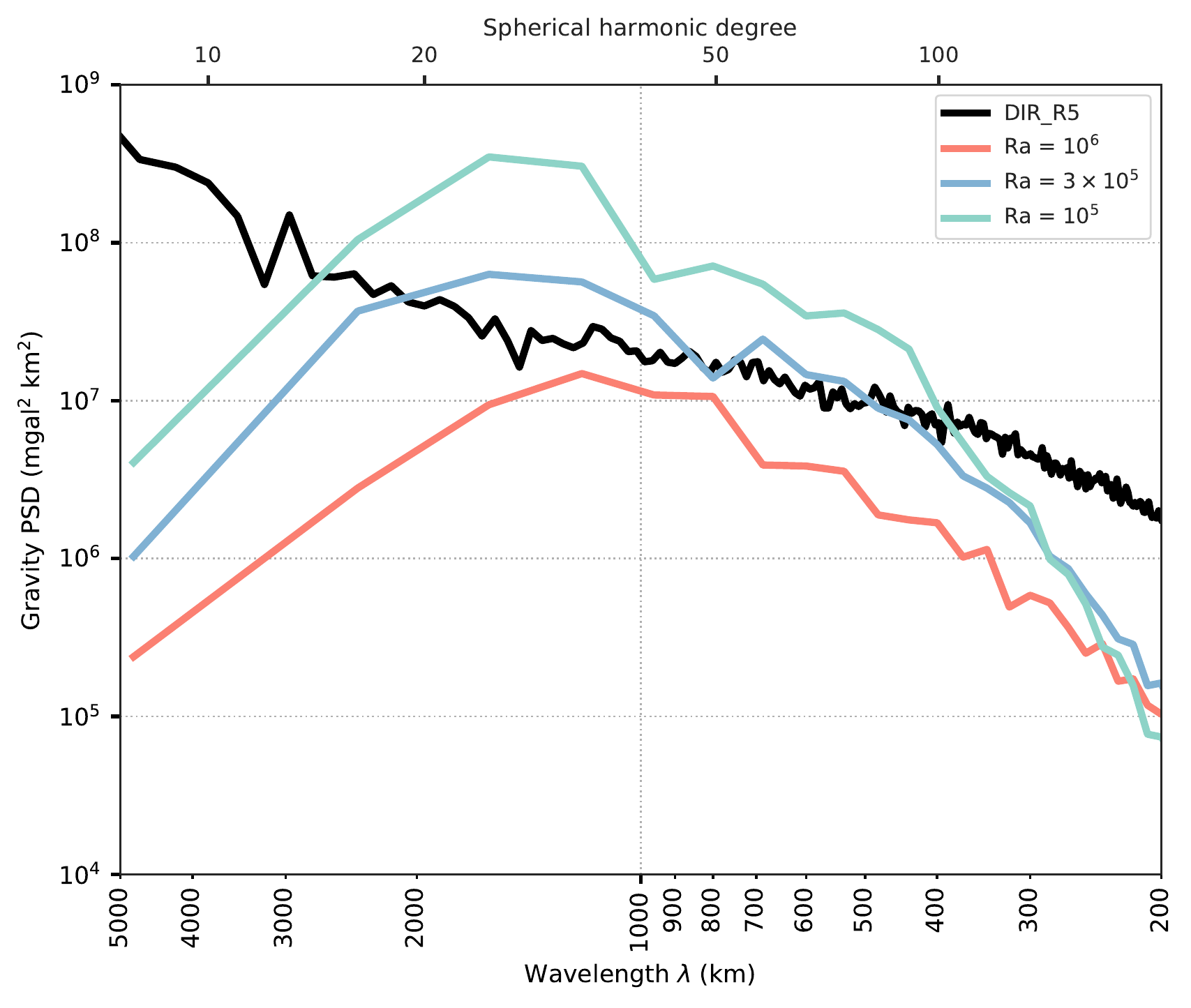}
\caption{Power spectral density of gravity anomalies at the top of the 
convecting box as in 
\autoref{fig:grav_spec}(a), but showing the variation with Rayleigh number for 
the 
free-rigid simulations.}
\label{fig:grav_spec_Ra}
\end{figure}

The behaviour of the power spectral density with Rayleigh number is 
illustrated in \autoref{fig:grav_spec_Ra} for the free-rigid simulations. The 
principal effect of changing Rayleigh number is to change the amplitude of the 
power spectral density. This can be understood from the scalings of RMS gravity 
with Rayleigh number above, and the Parseval's theorem result
\begin{linenomath}
\begin{equation}
 \Delta g_\text{RMS}^2 = \frac{1}{2 \pi} \int \phi_\text{PSD}^{\Delta g} (k)\, 
k \; \d k
\end{equation}
\end{linenomath}
which relates the square of the RMS value $\Delta g_\text{RMS}$ to its power 
spectral density $\phi_\text{PSD}^{\Delta g}(k)$. Since RMS gravity in 
\autoref{tab:params2} scales roughly as $\text{Ra}^{-0.6}$, the power spectral 
density would be expected to scale as $\text{Ra}^{-1.2}$. This effect can be 
seen in \autoref{fig:grav_spec_Ra}: reducing the Rayleigh number by a factor of 
10 leads to a little over a order of magnitude shift in the amplitude of the 
spectra. In addition to the change in amplitude, there are some more subtle 
changes in the spectra associated with changing the Rayleigh number. The lower 
Rayleigh number simulations appear to have relatively higher power at long 
wavelength than short wavelengths. This is as expected from the nature of the 
boundary layers, which are thicker for the lower Rayleigh number runs.

\subsection{The relationship between temperature and potential 
temperature}\label{sec:ptemp}

To perform melting calculations it is necessary to convert from dimensionless 
potential temperature back to real temperature. In this section we describe 
this conversion, and justify the approximate form of energy conservation that 
has been used in \eqref{eq:dimless_energy_cons}.
In dimensional variables, conservation of energy can be written
\begin{linenomath}
\begin{equation}
\rho T \frac{\D S}{\D t} = k \nabla^2 T + \Psi \label{eq:energy}
\end{equation}
\end{linenomath}
where $S$ is the specific entropy of a fluid parcel, $k$ is the thermal 
conductivity (assumed constant), and $\Psi$ is the viscous dissipation. The 
potential temperature $\theta$ can be defined in differential form as
\begin{linenomath}
\begin{equation}
 \d S = c_p \d \left(\log \theta \right)
\end{equation}
\end{linenomath}
where $c_p$ is the specific heat capacity at constant pressure (also assumed 
constant). The energy equation \eqref{eq:energy} becomes
\begin{linenomath}
\begin{equation}
 \frac{\D \theta}{\D t} =  \kappa \frac{\theta}{T} \nabla^2 T,
\end{equation}
\end{linenomath}
where $\kappa = k / (\rho c_p)$ is the thermal diffusivity, and viscous 
dissipation has been neglected. The use of 
\eqref{eq:dimless_energy_cons} as a dimensionless governing equation for 
potential temperature is justified provided the approximation
\begin{linenomath}
\begin{equation}
\frac{1}{\theta} \nabla^2 \theta \approx \frac{1}{T} \nabla^2 T 
\label{eq:keyapprox}
\end{equation}
\end{linenomath}
is accurate, and that viscous dissipation is sufficiently small to be neglected.

\subsubsection{No melting}\label{sec:nomelt}

In the absence of melting, the differential of potential temperature can be 
related to those of temperature and pressure through the standard relationship
\begin{linenomath}
\begin{equation}
 c_p \frac{\d \theta}{\theta} = \d S = c_p \frac{\d T}{T} - \frac{\alpha}{\rho} 
\d P \label{eq:differential}
\end{equation}
\end{linenomath}
where $\alpha$ is the thermal expansivity. The principal variation in pressure 
is hydrostatic. Writing $\d P = - \rho g\, \d z$, \eqref{eq:differential} can 
be written in terms of temperature and depth as
\begin{linenomath}
\begin{equation}
 \frac{\d \theta}{\theta} = \frac{\d T}{T} + \frac{\d z}{h_a} 
\label{eq:differential2}
\end{equation}
\end{linenomath}
where $h_a$ is the adiabatic scale height, defined by $h_a =  c_p / (\alpha 
g) \approx 3,300$ km. Integration of \eqref{eq:differential2} yields the 
relationship between temperature and potential temperature in regions which are 
not partially molten,
\begin{linenomath}
\begin{equation}
 T = \theta \exp \left(\frac{z_{\text{ref}}- z}{h_a} \right) 
\label{eq:potential_T_solid}
\end{equation}
\end{linenomath}
where $z_{\text{ref}}$ is a reference depth, the depth at which potential 
temperature is chosen to be equal to real temperature. This depth is chosen 
to be the Earth's surface in this work.

From \eqref{eq:differential2} it follows that
\begin{linenomath}
\begin{equation}
\frac{1}{\theta} \nabla^2 \theta = \frac{1}{T} \nabla^2 T + \frac{2}{h_a} 
\frac{1}{T} \pdd{T}{z} + \frac{1}{h_a^2}.
\label{eq:22}
\end{equation}
\end{linenomath}
The magnitude of the second and third terms on the right hand side relative to 
the first term scales approximately as $l/h_a$ and $(l/h_a)^2$ where $l$ is a 
typical scale over which the temperature varies. If that scale $l$ were the 
whole of the convecting layer then $l/h_a = d/h_a = 0.18$ (a parameter known 
as the Dissipation number), which is relatively small. In fact, the length 
scale of the vertical temperature variations will be much smaller than the 
layer depth, with boundary layer thicknesses on the order of 100 km or less, 
giving $l/h_a = 0.03$. Thus both the second and third terms on the right hand 
side of \eqref{eq:22} are sufficiently small that the approximation in 
\eqref{eq:keyapprox} is well justified in the regions that are not partially 
molten \citep{McKenzie1970}. The small Dissipation number for upper mantle 
convection also justifies the neglect of viscous dissipation term in 
\eqref{eq:energy}. 

An additional approximation has been made in writing the buoyancy term on 
the right-hand side of the Stokes equation in \protect{\eqref{eq:stokes}} in 
terms of the 
potential temperature $\theta$. Formally, density variations in the fluid are 
determined by the actual temperature, not potential temperature, and the 
right-hand side of \protect{\eqref{eq:stokes}} should be $-\rho_0 g \alpha T 
\hat{\mathbf{z}}$. The convective flow is driven by horizontal gradients in the 
actual temperature, not the potential temperature. From 
\protect{\eqref{eq:potential_T_solid}} it follows that the horizontal gradients 
in 
temperature are related to the horizontal gradients in potential temperature by
\begin{linenomath}
\begin{equation}
\pdd{T}{x} = \pdd{\theta}{x} \exp \left(\frac{z_{\text{ref}}- z}{h_a} \right), 
\end{equation}
\end{linenomath}
with a similar expression for the $y$-derivative. The horizontal gradients of 
potential temperature and actual temperature differ by an exponential factor 
whose magnitude is at most the exponential of the Dissipation number. For the 
upper mantle convection we consider here, this is a relatively small 
difference, and justifies the approximation made in using the potential 
temperature in \protect{\eqref{eq:stokes}}.

\subsubsection{Melting}

The convection simulations provide a 3D grid of potential temperature (entropy) 
within 
the box. To turn this into melting rate, the hydrous melting parametrisation of 
\citet{Katz2003} was used to calculate the expected degree of melting $F$ at 
each grid point assuming isentropic decompression melting to the given 
potential temperature and pressure at each grid point. The original 
parametrisation of degree of melting in \citet{Katz2003} is given with 
pressure and temperature as the thermodynamic variables. This parametrisation 
can be recast in terms of pressure and entropy (or potential temperature) by 
numerically integrating the relevant differential expressions. The differential 
expression for entropy when melting is
\begin{linenomath}
\begin{equation}
  c_p \frac{\d \theta}{\theta} = \d S = c_p \frac{\d T}{T} - 
\overline{\left(\frac{\alpha}{\rho}\right)} \d P + \Delta S \, \d F,
 \label{eq:m1}
\end{equation}
\end{linenomath}
where $F$ is the degree of melting, $\Delta S$ is the specific entropy 
difference between the two phases, and
\begin{linenomath}
\begin{equation}
 \overline{\left(\frac{\alpha}{\rho}\right)} = F \frac{\alpha_f}{\rho_f} + 
(1-F) \frac{\alpha_s}{\rho_s}. \label{eq:m2}
\end{equation}
\end{linenomath}
The \citet{Katz2003} parametrisation accounts for the different thermal 
expansivities $\alpha_s$, $\alpha_f$; and densities $\rho_s$, $\rho_f$ of 
the two phases (solid and liquid respectively), but the specific heat $c_p$ is 
assumed identical for both phases. All parameter values used here are identical 
to those in Table 2 of \citet{Katz2003}, with the exception of the specific 
entropy difference between the two phases which we set as $\Delta S = 
400$~J~kg$^{-1}$~K$^{-1}$.  The parametrisation provides the degree of melting 
$F$ as a function of temperature $T$ and pressure $P$ which can be expressed in 
differential form as
\begin{linenomath}
\begin{equation}
 \d F = \left(\pdd{F}{T}\right)_P \d T + \left(\pdd{F}{P}\right)_T \d P. 
\label{eq:m3}
\end{equation}
\end{linenomath}
Given a parcel of material that is subsolidus at a given potential temperature 
$\theta$ and depth $z$, \eqref{eq:potential_T_solid} gives the relationship 
between temperature and depth (or pressure) throughout the subsolidus region. 
Once the material crosses the solidus the relationship between temperature 
and pressure at constant entropy can be obtained by numerically 
integrating equations  \eqref{eq:m1}, \eqref{eq:m2}, and \eqref{eq:m3} with $\d 
S = 0$. Knowing the temperature and pressure then allows $F$ to be 
calculated. From a series of constant entropy integrations for different 
potential temperatures a parametrisation of $F$ as a function of entropy and 
pressure was generated. Using this entropy parametrisation, the grid of 
potential temperature and depth values in the box were converted to a grid of 
$F$ values. This grid of $F$ was then 
converted to a melting rate $\Gamma$ using
\protect \begin{equation}
 \Gamma = \pdd{F}{t} + \mathbf{v} \cdot \boldsymbol{\nabla} F
\end{equation}
where $\mathbf{v}$ is the velocity. The time derivative was calculated 
using a first-order accurate finite difference approximation. The 
spatial gradient was calculated using a 
second-order accurate finite difference approximation. The melting 
rate $\Gamma$ was vertically integrated to produce the plots shown in Figures 
\ref{fig:melt_gen} and \ref{fig:melt_gen_controls}. Only those regions 
where the melt rate was positive (i.e. melting) were included in the vertical 
integral. In the 
calculation of these figures, the advective term ($\mathbf{v} \cdot 
\boldsymbol{\nabla} F$) was larger by more than an order of a magnitude than 
the time dependent term ($\partial F 
/ \partial t$), and an excellent approximation to the melting rates can be 
obtained from the advective term alone. 

The calculation of melting rates were performed here as a postprocessing 
operation after running standard single-phase convection simulations. We 
assume that all the melt that is generated moves to the surface, 
and that none remains in the source region to freeze as the mantle material 
cools. It 
should be noted that this calculation neglects potentially important 
back-effects that melt can have on the flow, e.g. arising from the buoyancy of 
the 
melt, and the thermal effects of the consumption of latent heat. Indeed, in 
regions where melt is present, the approximation in \eqref{eq:keyapprox} can 
cease to be good approximation. However, the melting regions are only 
a small proportion of the overall domain, and the changes in temperature due to 
melting are small. For the free-rigid case with a lithospheric 
thickness of 80 km the average temperature change in the regions undergoing 
melting is $10^\circ$C, and the maximum change in the whole box is 
$90^\circ$C.  Furthermore such changes in temperature occur within the thermal 
boundary layer where heat is transported by conduction and where the 
temperature variations are large whether or not melting occurs. The effect 
of the temperature changes resulting from melting on the large-scale dynamics 
is therefore negligible.

There are other back-effects of melt extraction on the 
convective flow that have been neglected. When melt is extracted from the 
mantle, the remaining residue has 
a different density than it had before the melt was extracted. There is thus 
the 
potential for this depletion by melting to change buoyancy forces, and hence 
the flow. However, the density changes in the residue are small. Even for 20\% 
melt extraction the relative density changes on depletion are on the order of 
-0.5\% \protect{\citep{Schutt2006}}, equivalent to a density change from 
temperature variations of $125^\circ$C.

Another potential back-effect that has been neglected arises from the 
effect of volcanic loading on melt production rates. That changes in 
loading at the Earth's surface can influence melt production rates is 
well-known from studies that have looked at the volcanic response to changes 
in ice cover in Iceland \protect{\citep{Jull1996,Maclennan2002,Eksinchol2019}}. 
Beneath Iceland most melt generation occurs in the upper 100 km of the 
mantle, where the upwelling rate, of about 10 mm/a, is driven by the 
separating plates.  The thickness of the ice on Iceland reached about 3 
km, equivalent to a thickness of rock of about 1 km, over about $10^5$ years, 
corresponding to an equivalent rock accumulation rate of 10 mm/a.  Melt 
generation therefore ceases during the construction of the icecap, and 
all the melt that would normally have been generated during $10^5$ years 
is instead released when the ice melts, in about $10^3$ years.  The behaviour 
of melt generation within the convecting region beneath the lithosphere 
is very different. For the free-rigid case with a lithospheric 
thickness of 80 km shown in \autoref{fig:melt_gen}, the 
upwelling rate of the solid 
mantle where the melt generation rate is fastest is 27 mm/a.  The 
accumulation rate of melt at the surface is about 300 m/Ma, or 0.3 mm/a. 
This rate is therefore about 1/100th of the upwelling rate, and will 
have no significant effect on the melt generation rate. Moreover, volcanic 
loads at the surface will be eroded over time, and their influence on the 
mantle beneath is further attenuated by the finite elastic strength of the 
overlying lithosphere.

\subsection*{Acknowledgments}

The basis of this paper is Matthew Lees's Part III project in Earth 
Sciences. We thank Harro Schmeling and an anonymous reviewer for their 
comments which improved the paper. We thank Julianne 
Dannberg and Mark Hoggard for 
their advice.  We thank the Leverhulme Trust for support. This work was 
performed using resources provided by the Cambridge Service for Data Driven 
Discovery (CSD3) operated by the University of Cambridge Research Computing 
Service (\url{www.csd3.cam.ac.uk}), provided by Dell EMC and Intel using Tier-2 
funding from the Engineering and Physical Sciences Research Council (capital 
grant EP/P020259/1), and DiRAC funding from the Science and Technology 
Facilities Council (\url{www.dirac.ac.uk}). No new data has been generated in this 
work. Code for the numerical simulations of convection is available at 
\url{aspect.geodynamics.org}.

\makeatletter 
\renewcommand{\thefigure}{S\@arabic\c@figure}
\makeatother
\setcounter{figure}{0}
\begin{center}
\begin{figure}
\includegraphics[width=0.98\columnwidth]{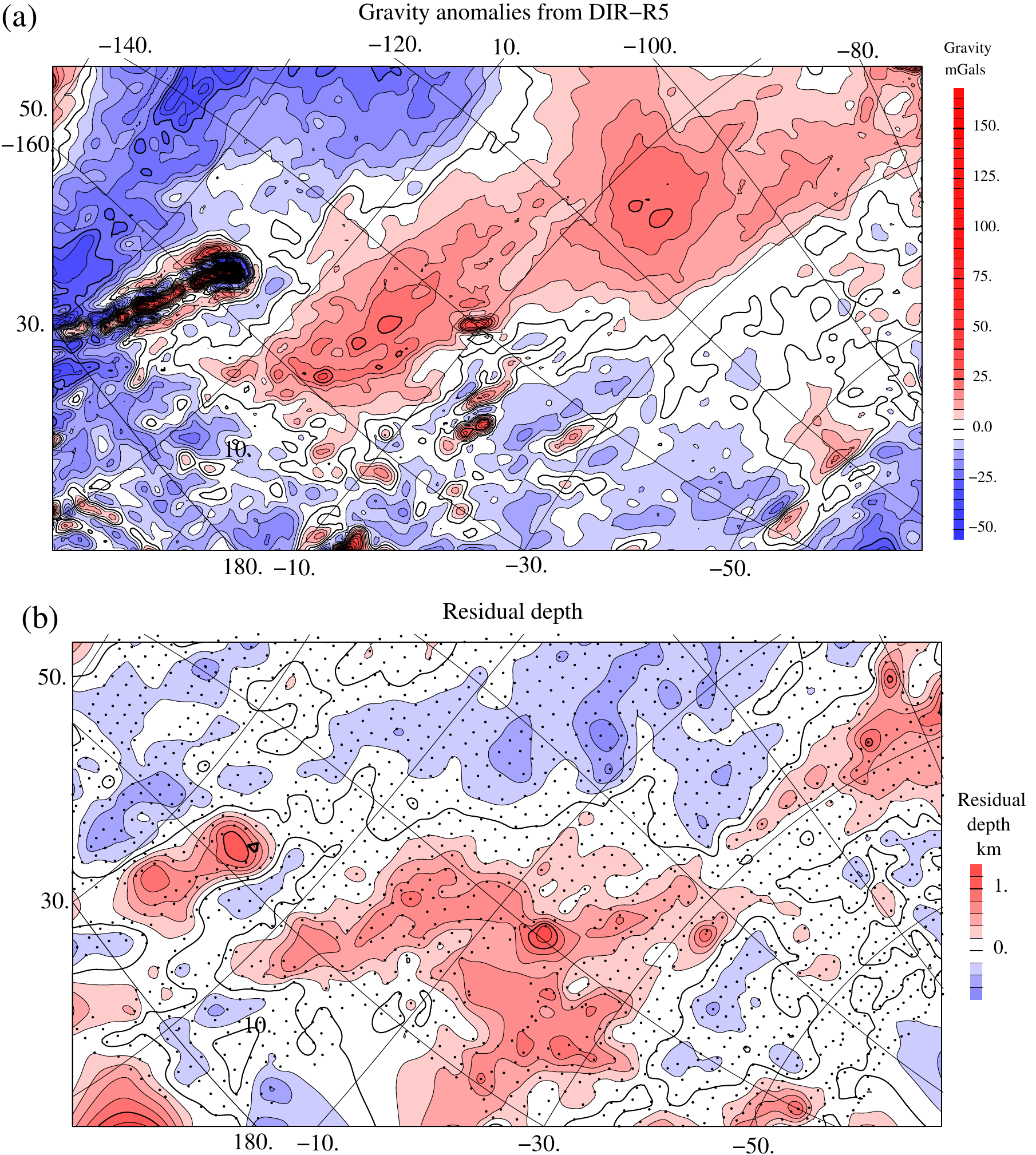}
\caption{(a) Gravity field for the Pacific from DIR-R5, with coefficients 
$l=2$ set to 0 and a filter applied, falling to 1/2 at 250 km, to remove the 
short wavelength components.  (b) Residual 
depths, averaged over $2^\circ \times 2^\circ$ boxes (Crosby et al. 2006). The 
dots show the locations of the resulting averages.   
Oblique Mercator projection with axis $40^\circ$N, $-50^\circ$E.} 
\end{figure}
\end{center}
\clearpage
\newpage
\begin{center}
\begin{figure}
\includegraphics[width=0.99\columnwidth]{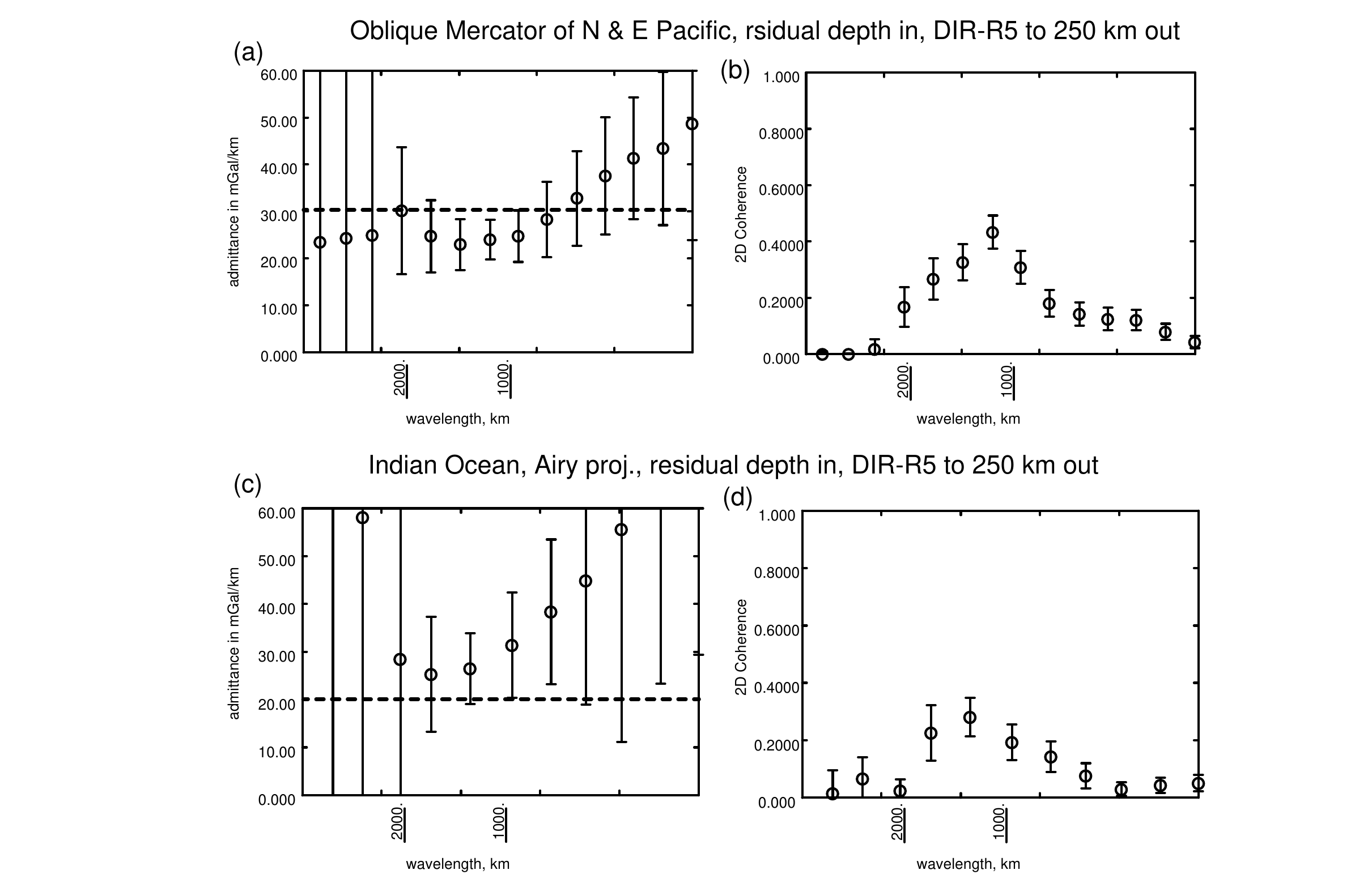}
\caption{Admittance and coherence from the same regions used for Figure 12 
(c)-(f), calculated using Hoggard et al.'s (2017) estimates of residual depth 
rather than those of Crosby et al. (2006).} 
\end{figure}
\end{center}
\begin{center}
\begin{figure}
\includegraphics[width=0.86\columnwidth]{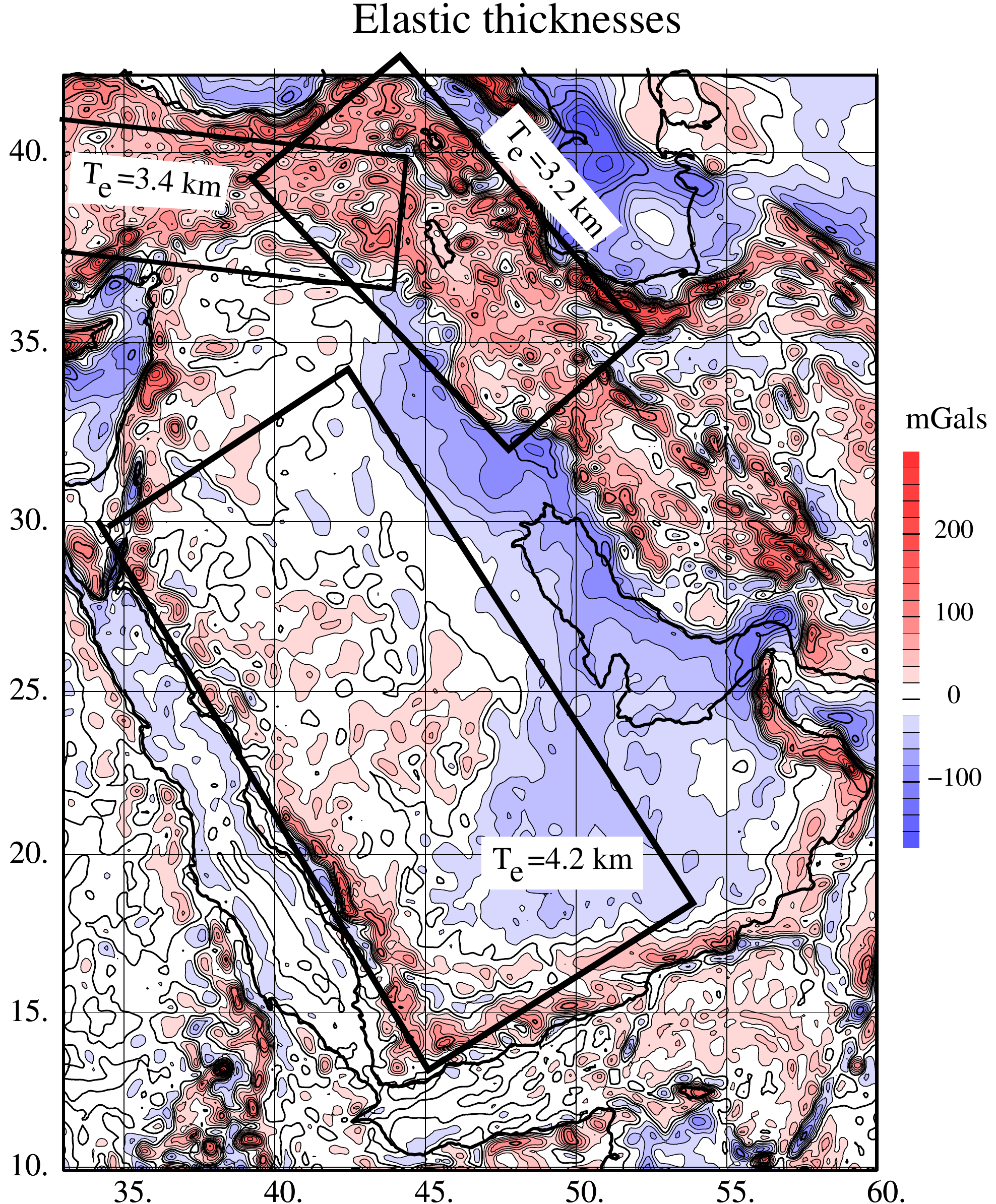}
\caption{Boxes used to estimate the elastic thickness $T_e$ of different parts 
of the Middle East, superimposed on gravity anomalies from Eigen6c (F\"{o}rste 
et al. 2011), with the coefficients from $l=2$ to 7 to 0, and  applying a 
taper $f,=(l-7)/5$, to those from $l=8$ to 11.  A low pass filter falling to 
1/2 at 50 km was applied to the coefficients to remove the short wavelength 
anomalies from uncompensated topography.} 
\end{figure}
\end{center}
\begin{center}
\begin{figure}
\includegraphics[width=0.99\columnwidth]{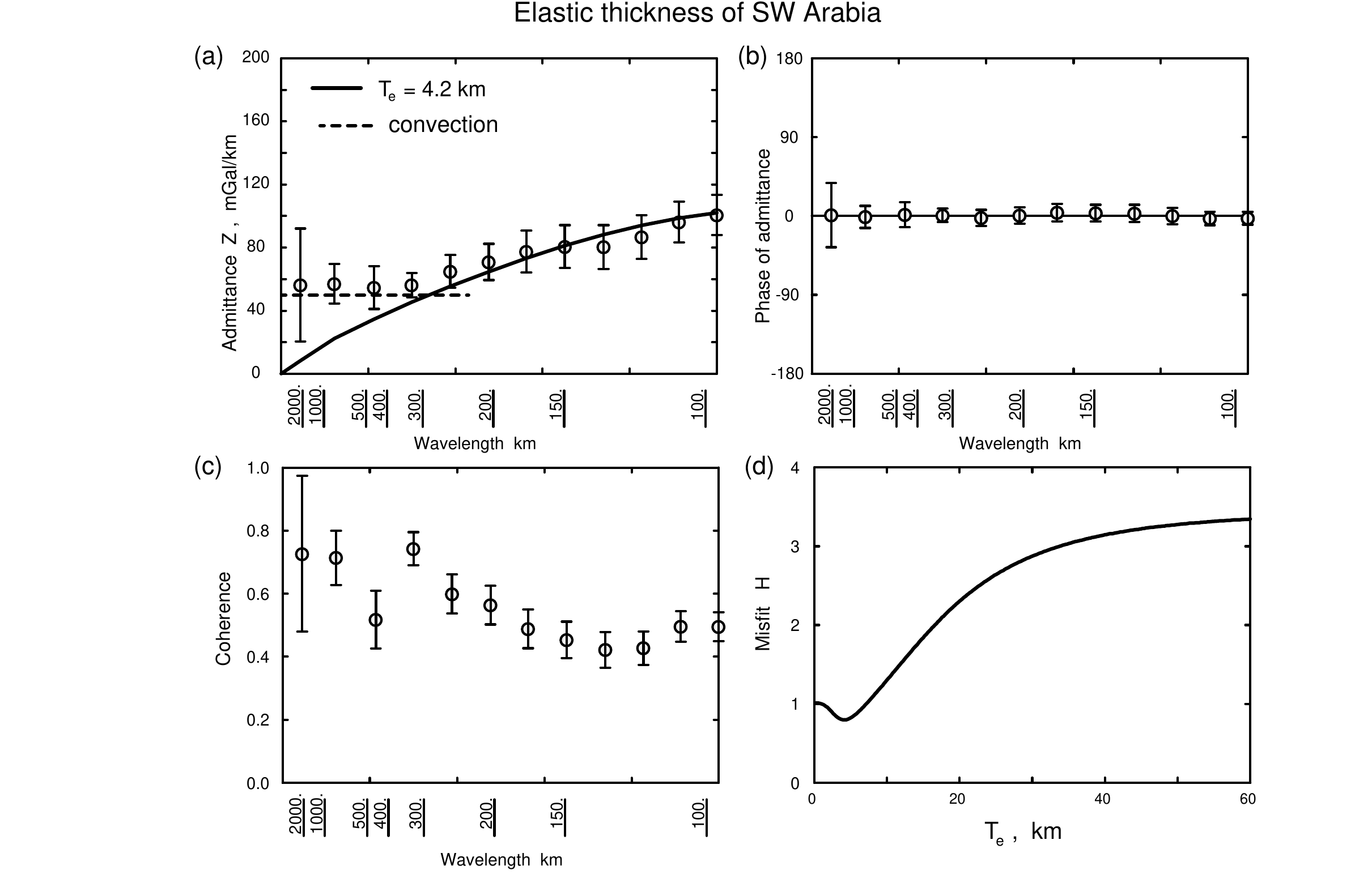}
\caption{Estimate of the elastic thickness for S and W Arabia (see Figure S3) 
from
the admittance, taking the topography as input, gravity from Eigen6c as 
output.} 
\end{figure}
\end{center}
\clearpage
\newpage
\begin{center}
\begin{figure}
\includegraphics[width=0.99\columnwidth]{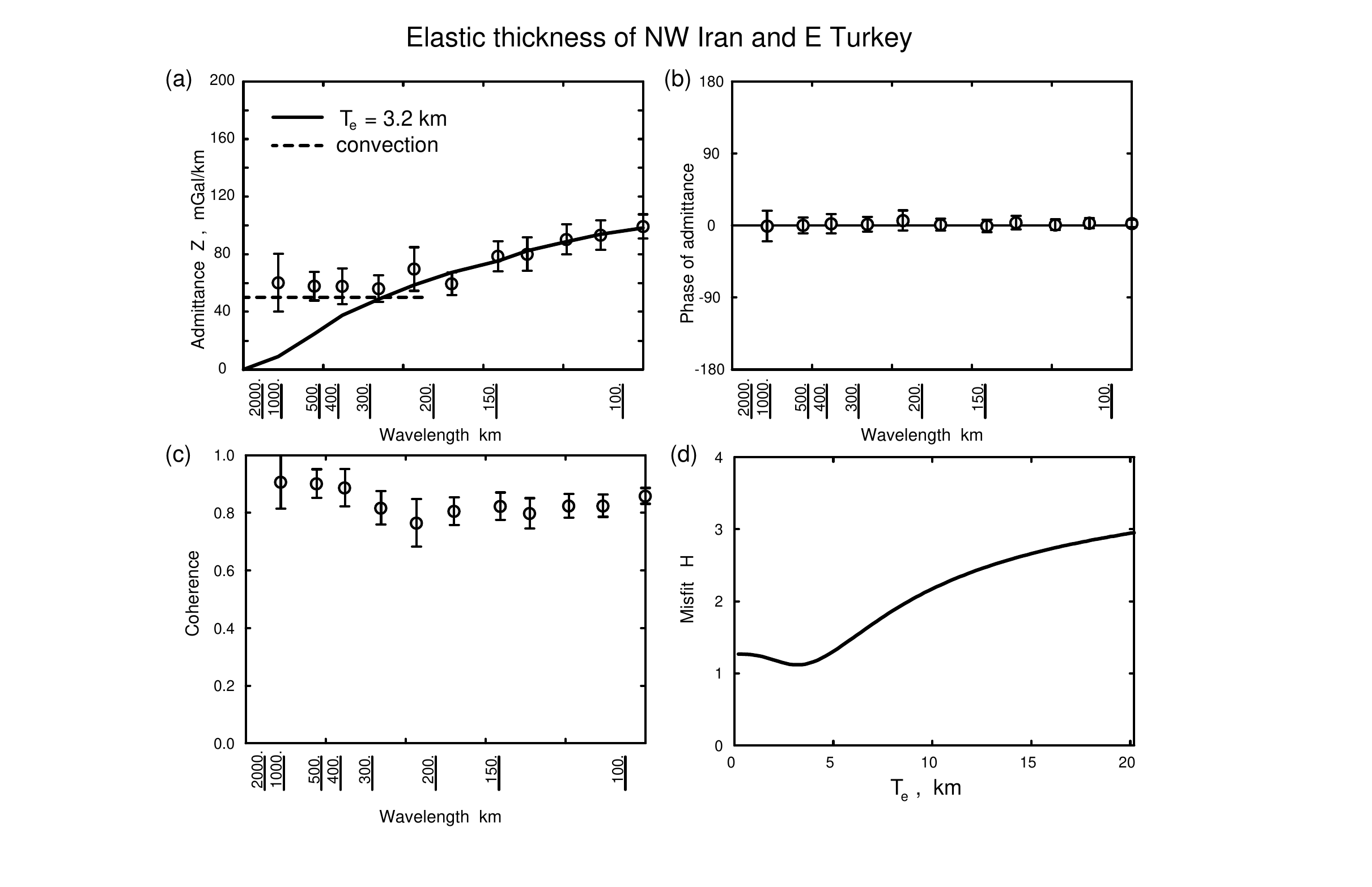}
\caption{Estimate of the elastic thickness for E Turkey and NW Iran 
(see Figure S3) from
the admittance, taking the topography as input, gravity from Eigen6c as 
output.} 
\end{figure}
\end{center}
\clearpage
\newpage
\begin{center}
\begin{figure}
\includegraphics[width=0.99\columnwidth]{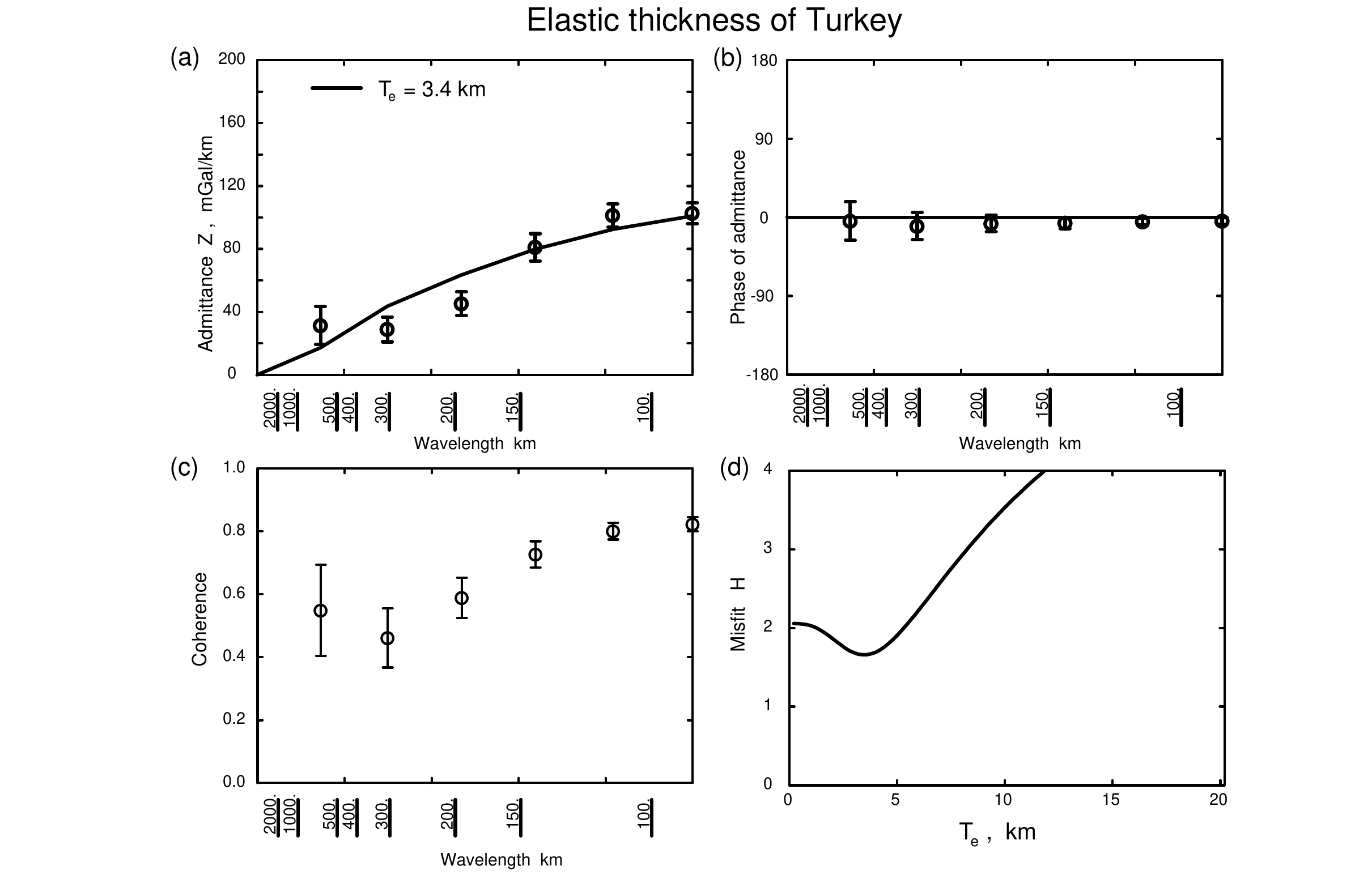}
\caption{Estimate of the elastic thickness for Anatolia (see Figure S3) from
the admittance, taking the topography as input, gravity from Eigen6c as 
output.} 
\end{figure}
\end{center}

\end{document}